\documentclass[apj]{emulateapj}
\usepackage{times}

\shorttitle{AGN radio luminosity and cluster environment}
\shortauthors{Ineson et al.}
\begin{document}
\title{Radio-loud AGN: is there a link between luminosity and cluster environment?}
\author{J. Ineson\altaffilmark{1}, J. H. Croston\altaffilmark{1}, M. J. Hardcastle\altaffilmark{2}, R. P. Kraft\altaffilmark{3}, D. A. Evans\altaffilmark{3}, M. Jarvis\altaffilmark{4}\altaffilmark{2}\altaffilmark{5}}
\altaffiltext{1}{School of Physics and Astronomy, University of Southampton,
  Southampton SO17 1BJ, UK.; J.Croston@soton.ac.uk}
\altaffiltext{2}{School of Physics, Astronomy and Mathematics, University of
  Hertfordshire, Hatfield, Hertfordshire AL10 9AB, UK}
\altaffiltext{3}{Harvard-Smithsonian Center for Astrophysics, 60 Garden Street, Cambridge, MA 02138, USA}
\altaffiltext{4}{Astrophysics, Department of Physics, Keble Road, Oxford OX1 3RH, UK}
\altaffiltext{5}{Department of Physics, University of the Western Cape, Private Bag
X17, Bellville 7535, South Africa}

\begin{abstract}

We present here the first results from the Chandra ERA (Environments of Radio-loud AGN) Large Project, characterizing the cluster environments of a sample of 26 radio-loud AGN at $z \sim 0.5$ that covers three decades of radio luminosity. This is the first systematic X-ray environmental study at a single epoch, and has allowed us to examine the relationship between radio luminosity and cluster environment without the problems of Malmquist bias. We have found a weak correlation between radio luminosity and host cluster X-ray luminosity, as well as tentative evidence that this correlation is driven by the subpopulation of low-excitation radio galaxies, with high-excitation radio galaxies showing no significant correlation. The considerable scatter in the environments may be indicative of complex relationships not currently included in feedback models.
\end{abstract}



\section{INTRODUCTION}\label{sec:intro}

Understanding how the properties of radio-loud AGN relate to their cluster environments is crucial for our understanding of the role of AGN feedback in galaxy evolution. Suppression of star formation by feedback from this type of AGN is now an important feature of simulations of galaxy evolution, allowing them to match the observed galaxy sizes and star formation rates (Croton et al. 2006). The fact that the radio jets disturb the cluster environment can be seen in detailed studies of nearby radio galaxies (eg Kraft et al. 2003; Fabian et al. 2003; Forman et al. 2005; Croston et al. 2009). The processes involved are described in the review by McNamara \& Nulsen (2007). Galaxy feedback via radio jets is however a complex, two-way process since the environment is also expected to affect the properties of the radio galaxy, and the relationship between energy input from the radio-loud AGN and environment, and how their relationship evolves with epoch is as yet poorly understood. The radio jets transport energy a considerable distance into the cluster and are themselves modified by the intra-cluster medium. Do the properties of the large-scale cluster environment in their turn affect the feedback loop maintaining the AGN, or are the AGN properties determined by the more local environment of the host galaxy? And how does does this disruption of the cluster environment affect its evolution? 

Thus two outstanding questions are whether the radio luminosity is related to the large-scale cluster environment, and whether typical environments evolve with epoch.

These questions have been in consideration for some time. At low redshifts, it has long been known that FRI galaxies appear to inhabit richer clusters than the more luminous FRII galaxies (eg Longair \& Seldner 1979; Prestage \& Peacock 1988). However, Prestage \& Peacock found that although there was a clear difference in average richness, there was also a large scatter in the richness of environments, with both types having examples at similar extremes. Hill \& Lilly (1991) extended this work by comparing the low redshift results with a flux-limited sample at $z \sim 0.5$. They found that at the higher redshift range, the FRII galaxies were spread more evenly over a wider range of cluster richnesses than at low redshift, raising the possibility that the environments of FRII galaxies evolve.

There are a number of selection biases that may have affected these early studies, and that still make sample selection difficult today. The overwhelming majority of radio galaxies in the local universe are low luminosity FRI galaxies. However, with flux-limited samples we see only increasingly luminous objects, predominantly FRIIs, as we increase redshift -- Figure~\ref{fig:Lr-z} (left) shows the problem clearly. This gives opposite biases in results at low and high redshifts and makes systematic comparisons between low and high redshift data difficult.

There is also a potential confounding factor in the choice of subsamples for comparisons. The studies cited above use the FR classification (Fanaroff \& Riley 1974), which is based on the location of the brightest area of luminosity in the radio lobes. Fanaroff \& Riley found that this corresponded to a reasonably clear division in radio luminosity, and this division was later found also to be related to optical luminosity (Owen \& Ledlow 1994). However, radio-loud AGN can also be classified by their optical spectral properties, ie whether or not the object has strong emission lines (eg Hine \& Longair 1979; Laing et al. 1994). High-excitation radio galaxies (HERGs), which have strong emission lines, incorporate both broad- and narrow-line radio galaxies (BLRGs and NLRGs), while low-excitation radio galaxies (LERGs) lack strong emission lines. The majority of HERGs are luminous FRII galaxies, whereas LERGs span the full range of radio luminosities, including almost all FRIs and a significant number of FRIIs. Hine \& Longair, for example, found that about 10\% of weak radio galaxies were HERGs, with the percentage increasing to over 70\% at high radio luminosities. Willott et al. (2001) also found that the proportion of HERGs increased with radio luminosity, with examples of HERGs and LERGs at all luminosities. Thus subpopulations based on the Fanaroff-Riley classification will consist of mainly, but not exclusively, LERGs for the FRI class and a mixture of HERGs and LERGs for the FRII class.

Discussions of AGN unification (eg Antonucci 1993; Laing et al. 1994) recognized that although radio-loud QSOs, BLRGs and NLRGs could be interpreted as the same class of object viewed at different orientations, LERGs appeared to be a different class of object. Further differences between the two classes emerged (eg Whysong \& Antonucci 2004; Ogle et al. 2006; Chiaberge et al. 2002; Evans et al. 2006; Hardcastle et al. 2006) and it is now thought that HERGs are undergoing radiatively efficient accretion, while LERGs are radiatively inefficient (eg Hardcastle et al. 2007). The classes therefore appear to be fundamentally different.

More recent studies into the effects of cluster richness therefore tend to split samples by spectral class rather than, or as well as, FR class. Looking at relatively low redshift samples ($z<0.4$), Best (2004) and Hardcastle (2004) both found that the different types of HERG all inhabit poor environments, but that LERGs are spread across a broader range of environments. Furthermore, the environments of the different types of HERG were all similar, as would be expected if they were different aspects of the same class of object. Gendre et al. (2013) obtained the same results for HERGs and LERGs for $z<0.3$, and also found that the FR class was independent of excitation class. At higher redshifts, Harvanek et al. (2001) found QSOs at $0.4<z<0.65$ in richer environments than at $z<0.4$, and Belsole et al. (2007), using a sample of FRII galaxies that were mostly HERGs ($0.45<z<1.0$), also found them inhabiting relatively rich environments. Since HERGs are mostly FRII galaxies and LERGs are both FRI and FRIIs, these results are compatible with Hill \& Lilly's results for FRI and FRII galaxies.

Does this imply that the environments of HERGs change with time, or, given the combination of the Malmquist bias and the paucity of high luminosity radio galaxies at low redshifts, does this imply that the environments of high luminosity sources are typically richer than those of low luminosity sources? Best (2004), using a sample with $z<0.1$, found a strong correlation between radio luminosity and environment richness for LERGs, but not for HERGs. Belsole et al. (2007) found no correlation for their high-$z$ FRII sample, but Wold et al. (2000; $0.5<z<0.9$) and Falder et al. (2010; $z \sim 1.0$) both found a correlation for high redshift, radio-loud QSOs. There is therefore some evidence that radio luminosity is related to the richness of the cluster environment for at least some classes of radio galaxy, but this does not exclude the possibility of evolution with epoch. However, Wold et al. (2000) and McLure \& Dunlop (2001) both compared compared their samples with results from studies at different redshifts and found no evidence of a variation of environment with redshift -- evidence supporting a link between environment and radio luminosity rather than epoch. So the picture of the relationship between radio-loud AGN and their large-scale cluster environments, and for the evolution of those environments, remains confused, and the long-term aim of our research program is to clarify these issues.

The studies described above use a variety of measures of cluster richness, raising questions about the compatibility of their results, and how well the total cluster mass is being traced. The scaling relations between mass, luminosity and temperature of the ICM are well defined for galaxy clusters (eg Arnaud et al. 2007; Vikhlinen et al. 2006; Pratt et al. 2009), and so the X-ray luminosity of the ICM is one of the most well understood measures of cluster richness. Despite the scatter in the mass-luminosity relationship, it should provide a more accurate proxy for cluster mass than the optical measures. There is evidence that it is strongly related to the optical measures; for example, Yee \& Ellingson (2003) have found a correlation between the galaxy cluster center correlation amplitude B$_{gc}$ and X-ray luminosity, and Ledlow et al. (2003) between Abell class and X-ray luminosity. X-ray observations also have the advantage of directly probing the medium into which the radio jets are propagating, and so the relationship between X-ray cluster properties and radio-source characteristics might be expected to be tighter than for measures associated with the cluster galaxy population. 

In this paper, we present the first results of the {\it Chandra} ERA (Environments of Radio-loud AGN) program, whose aim is to make a systematic, X-ray based examination of the effects of cluster environment and epoch on the properties of radio-loud AGN for the first time. If the relationship between these properties can be clearly disentangled, then well-characterized relationships could be used for modeling radio-jet feedback in galaxy evolution models. In order to separate the effect of luminosity from that of redshift, we use a sample limited to a narrow redshift range but spanning a wide range of radio luminosities. This paper contains the results of this first part of this program. In future papers, we will compare these results with observations at different epochs to look for evidence of environmental evolution with redshift.

Throughout this paper we use a cosmology in which $H_0 = 70$ km s$^{-1}$ Mpc$^{-1}$, $\Omega_{\rm m} = 0.3$ and $\Omega_\Lambda = 0.7$. Unless otherwise stated, errors are quoted at the 1$\sigma$ level.

\section{THE SAMPLE}

We made use of the sample of McLure et al. (2004), which contains all narrow-line and low-excitation radio galaxies with redshifts between 0.4 and 0.6 from four flux-limited, spectroscopically complete, low-frequency radio surveys of the northern hemisphere -- 3CRR (Laing et al. 1983), 6CE (Eales et al. 1997; Rawlings et al. 2001), 7CRS (Lacy et al. 1999; Willott et al. 2003), TexOx-1000 (Hill \& Rawlings 2003). This sample is ideal for achieving our aim of comparing radio luminosity and environment richness without contamination by evolution, as it covers three decades in radio luminosity in a small redshift range while being distant enough to contain high luminosity radio galaxies but near enough for low luminosity galaxies still to be detectable and for X-ray observations of the ICM to be feasible with reasonable exposure times.

\begin{figure*}
  \hfill
  \begin{minipage}[t]{.5\textwidth}
    \begin{center}  
      \includegraphics[scale=0.23]{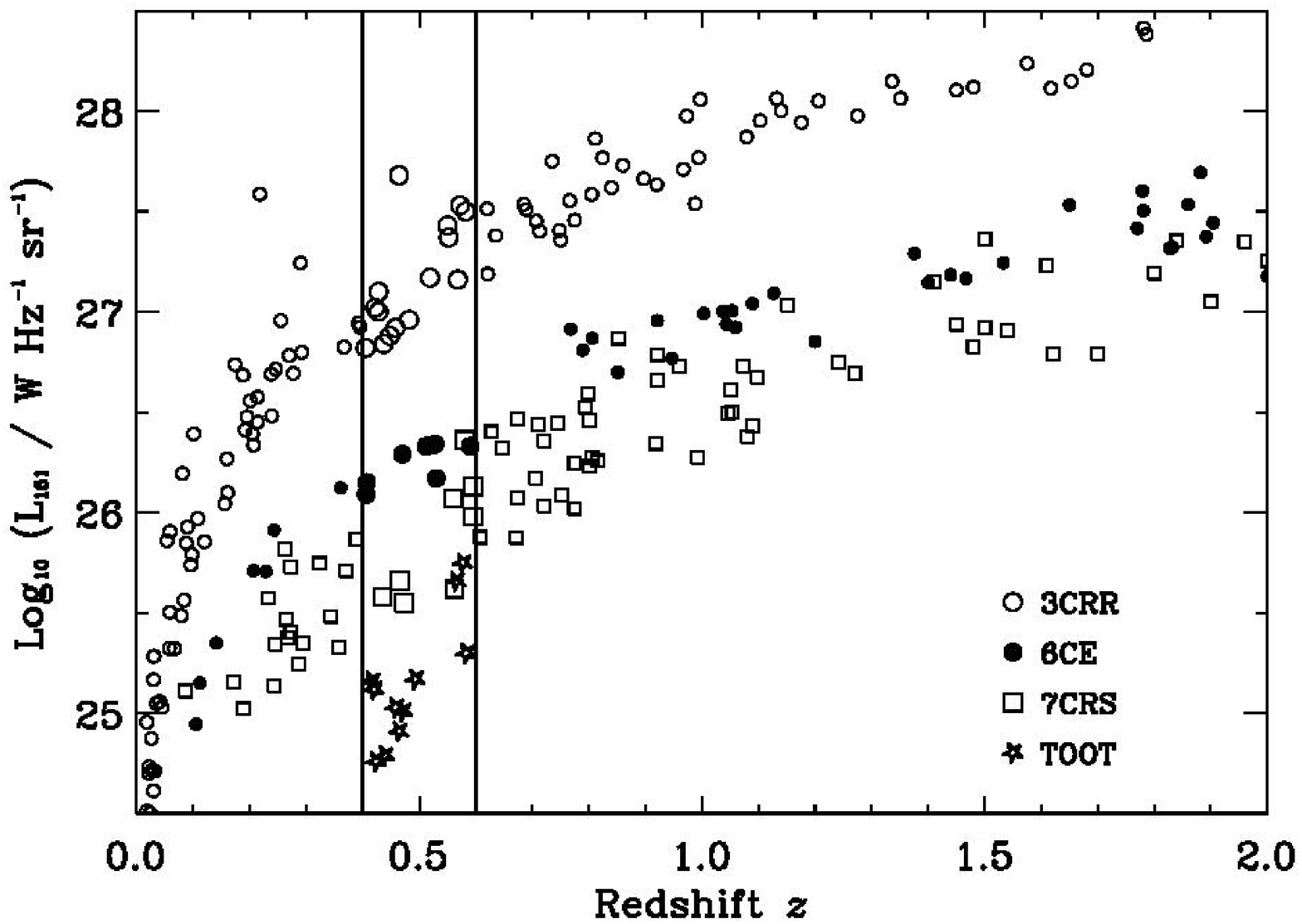}
    \end{center}
  \end{minipage}
  \hfill
  \begin{minipage}[t]{.45\textwidth}
    \begin{center}  
      \includegraphics[scale=0.36]{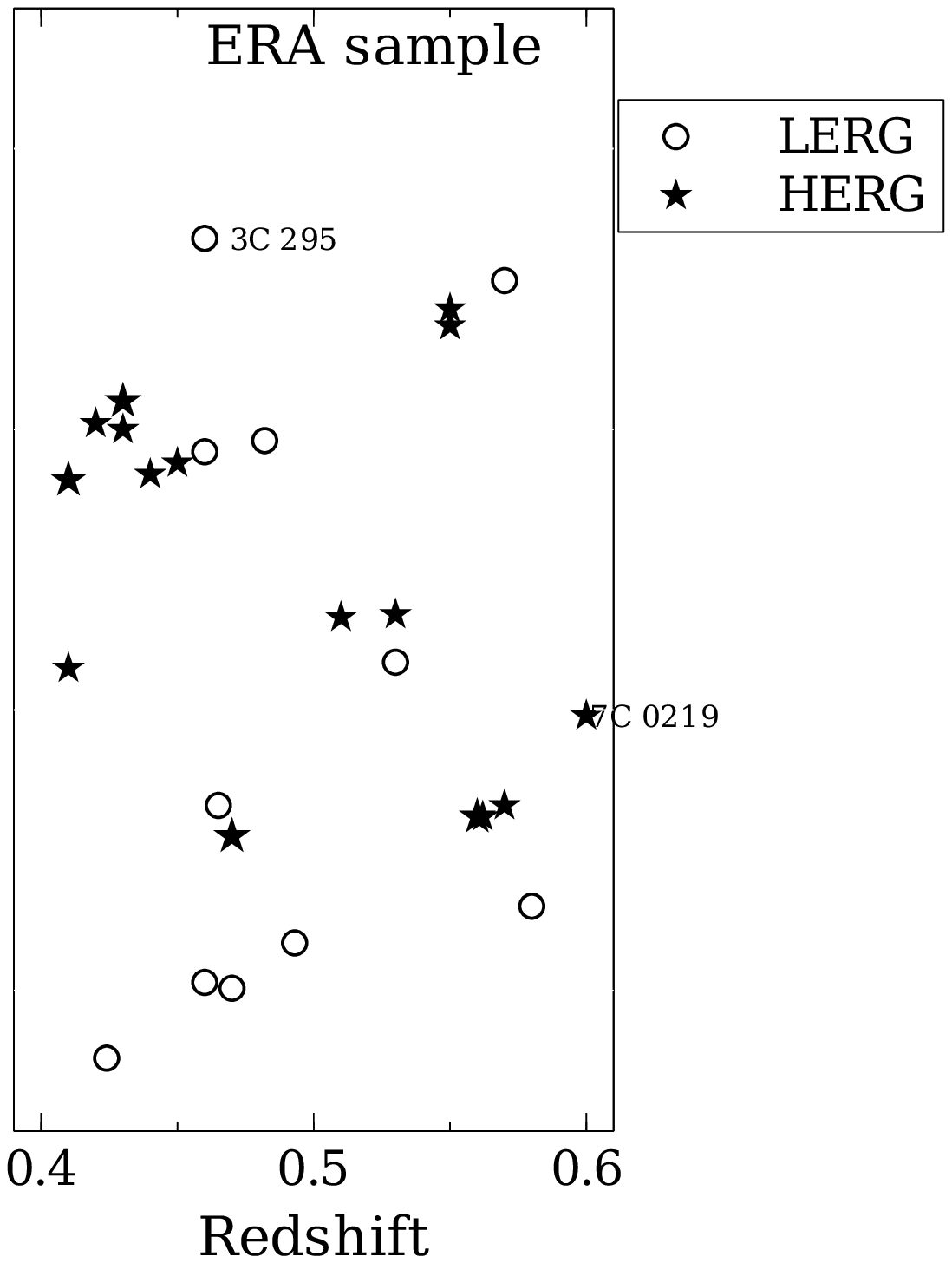}
    \end{center}
  \end{minipage}
  \hfill
\caption{On the left, 151-MHz radio luminosity vs redshift for narrow-line radio galaxies in the 3CRR, 6CE and 7CRS samples, with galaxies from the TOOT survey added in the range $0.4<z<0.6$ (McLure et al. 2004). The McLure et al. sample is in the $0.4<z<0.6$ redshift interval. On the right, the ERA subsample used in this project.}
\label{fig:Lr-z}
\end{figure*}

\begin{figure*}
\plottwo{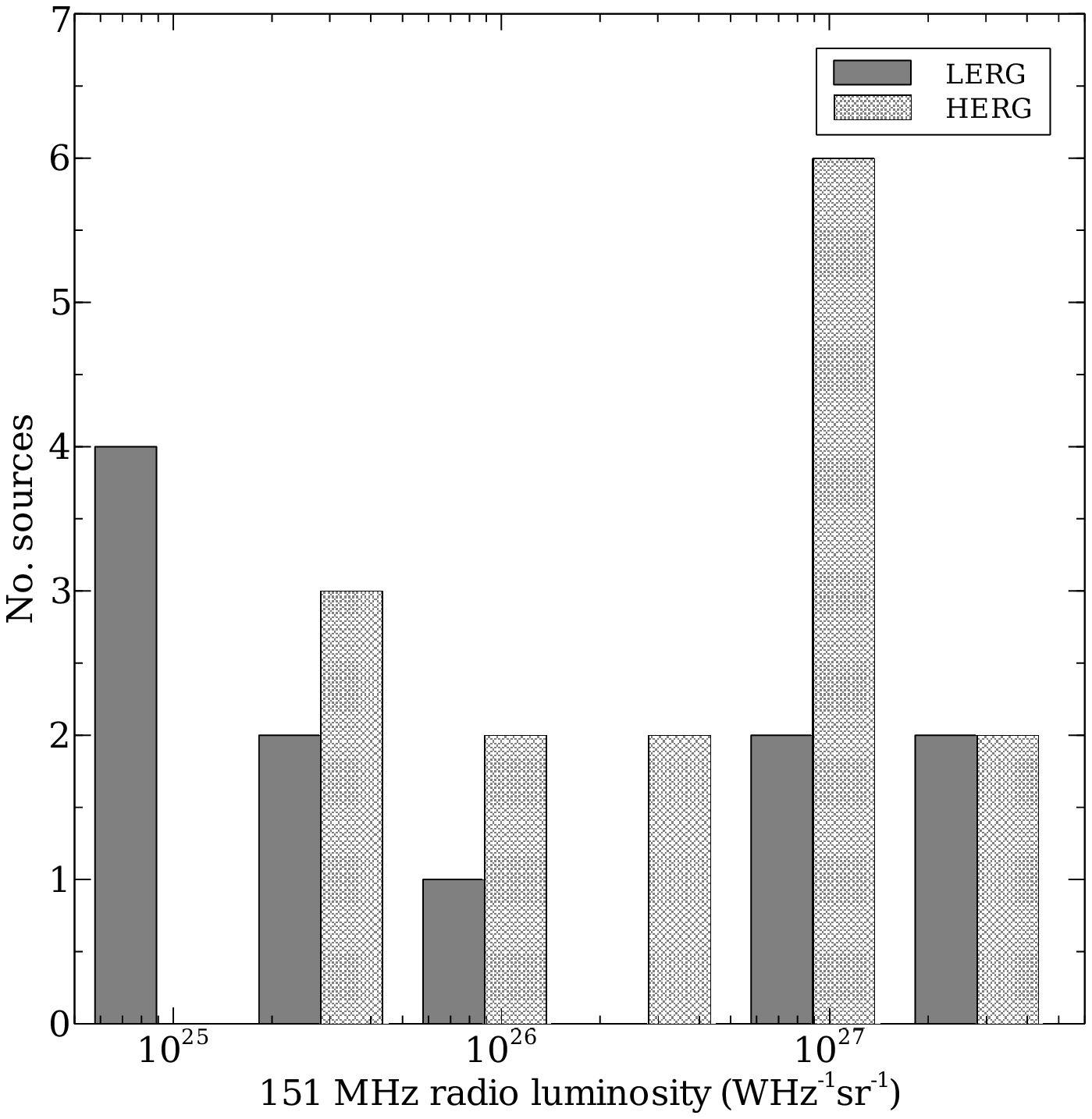}{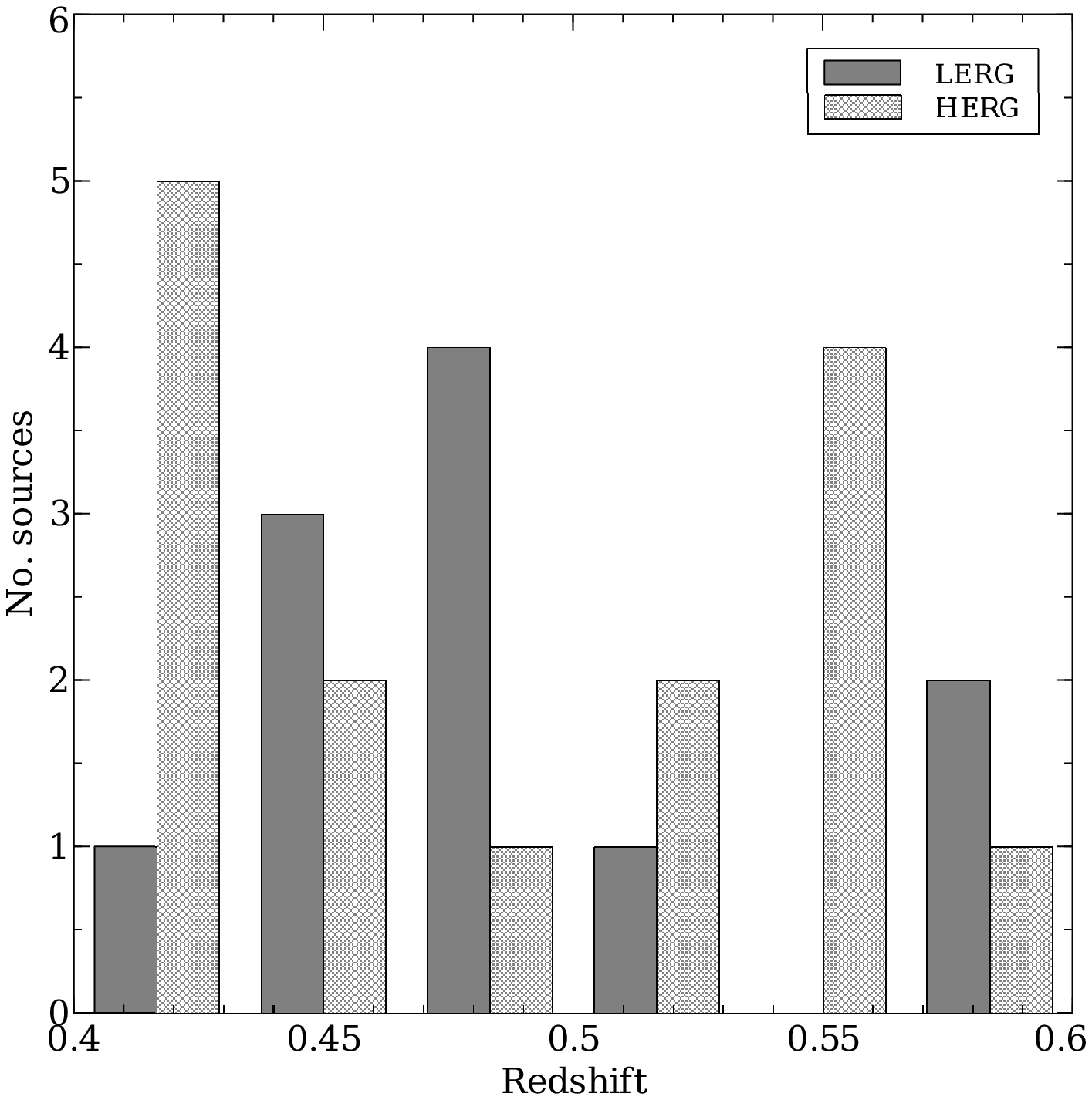}
\caption{The ERA sample; Radio luminosity coverage (left) and redshift coverage (right).}
\label{fig:ERA-histos}\end{figure*}

The luminosity and redshift range for the four surveys are shown on the left of Figure~\ref{fig:Lr-z}, with the McLure et al. (2004) sample highlighted. The sample contains 41 sources covering three decades of radio luminosity, and includes high- and low-excitation sources and a range of radio source morphologies. Host galaxy properties for the full sample were derived by McLure et al. (2004) from HST WFPC2 observations, and an optical environment study has recent been carried out by Herbert et al. (2013), using the spatial covariance function $B_{\rm gq}$. Because of the amount of X-ray observing time required, we could not use the entire McLure et al. sample for this project. We therefore constructed the ERA sample, a representative subsample of 24 sources from the McLure et al. sample covering its full range of radio luminosity and containing the same subgroups. We limited our sample to sources with extended lobes, so compact sources were excluded from our selection. Ten of the sources had already been imaged in X-ray by {\it XMM-Newton} or {\it Chandra}, and we obtained new {\it Chandra} and {\it XMM-Newton} observations of fourteen sources for this project. Subsequently, X-ray data for another two sources from the parent sample have become available and they have been included. Figure~\ref{fig:Lr-z} shows radio luminosity plotted against redshift for the ERA sample (right) alongside the McLure et al. (2004) sample (left), and Figure~\ref{fig:ERA-histos} shows the coverage of the ERA sample.

Table~\ref{tab:rawdata} lists details of the 26 sources. Positions and redshifts were taken from McLure et al. (2004). For all galaxies except 3C~457, we obtained Galactic column densities from Dickey \& Lockman (1990) via the \textsc{heasarc} tools; for 3C~457, we used the higher column density found by Konar et al. (2009). Excitation type was taken from McLure et al. (2004), except for 3C~295. 3C~295 is classified as a LERG in the on-line 3CRR catalog\footnote{\url{http://3crr.extragalactic.info/}} based on the results of Lawrence et al. (1996), but this classification is questioned (Varano et al. 2004). We have here classified 3C~295 as a LERG, but where appropriate have analyzed LERG subsamples with and without 3C~295. 7C~0219+3423's classification is also uncertain; we have followed McLure et al. in classifying it as a possible HERG.


\renewcommand{\arraystretch}{1.0}
\begin{deluxetable*}{lllcccccc}
\tablecolumns{9}
\tablewidth{0pc}
\tablecaption{The ERA sample, in order of radio luminosity}
\tablehead{
	\colhead{Source}&
	\multicolumn{2}{c}{RA  (J2000)   Dec}               &
	\colhead{Redshift}&
	\colhead{Scale}&
	\colhead{log$_{10}$ L$_{151}$}&
	\colhead{Radio}&
	\colhead{Type}&
	\colhead{nH}\\
	\colhead{}                                          &        
	\colhead{{h}\phn{m}\phn{s}}                         &
	\colhead{\phn{\arcdeg}~\phn{\arcmin}~\phn{\arcsec}} &
	\colhead{}&
	\colhead{kpc arcsec$^{-1}$}&
	\colhead{WHz$^{-1}$sr$^{-1}$}&
	\colhead{structure\tablenotemark{a}}&
	\colhead{}&
	\colhead{x10$^{20}$ cm$^{-2}$}\\}
\startdata
TOOT 1301+3658&13 01 25.03&+36 58 09.4&0.424&5.56&24.76&J&LERG&1.32\\
TOOT 1255+3556&12 55 55.83&+35 56 35.8&0.471&5.91&25.01&J&LERG&1.30\\
TOOT 1626+4523&16 26 48.5&+45 23 42.6&0.458&5.82&25.03&FD&LERG&1.12\\
TOOT 1630+4534&16 30 32.8&+45 34 26.0&0.493&6.06&25.17&J&LERG&1.26\\
TOOT 1307+3639&13 07 27.07&+36 39 16.4&0.583&6.60&25.30&J&LERG&1.14\\
7C 0223+3415&02 23 47.24&+34 15 11.9&0.473&5.92&25.55&CD&HERG&6.31\\
7C 1731+6638&17 31 43.84&+66 38 56.7&0.562&6.48&25.62&CD&HERG&3.84\\
7C 0213+3418&02 13 28.39&+34 18 30.6&0.465&5.87&25.66&CD&LERG&6.60\\
TOOT 1303+3334&13 03 10.29&+33 34 07.0&0.565&6.50&25.66&J&HERG&1.06\\
7C 0219+3423&02 19 37.83&+34 23 11.2&0.595&6.66&25.98&J&HERG?&6.30\\
6C 0850+3747&08 50 24.77&+37 47 09.1&0.407&5.43&26.15&FD&HERG&2.95\\
6C 1200+3416&12 00 53.34&+34 16 47.3&0.530&6.29&26.17&CD&LERG&1.62\\
6C 1132+3439&11 32 45.74&+34 39 36.2&0.512&6.18&26.33&CD&HERG&2.14\\
6C 0857+3945&08 57 43.56&+39 45 29.0&0.528&6.28&26.34&CD&HERG&2.64\\
3C 16&00 37 45.39&+13 20 09.6&0.405&5.41&26.82&CD&HERG&4.48\\
3C 46&01 35 28.47&+37 54 05.7&0.437&5.66&26.84&CD&HERG&5.66\\
3C 341&16 28 04.04&+27 41 39.3&0.448&5.74&26.88&CD&HERG&3.26\\
3C 200&08 27 25.38&+29 18 45.5&0.458&5.82&26.92&CD&LERG&3.74\\
3C 19&00 40 55.01&+33 10 07.3&0.482&5.98&26.96&CD&LERG&5.82\\
3C 457&23 12 07.57&+18 45 41.4&0.428&5.59&27.00&CD&HERG&22.3\tablenotemark{b}\\
3C 274.1&12 35 26.64&+21 20 34.7&0.422&5.55&27.02&CD&HERG&2.00\\
3C 244.1&10 33 33.97&+58 14 35.8&0.430&5.61&27.10&CD&HERG&0.58\\
3C 228&09 50 10.79&+14 20 00.9&0.552&6.42&27.37&CD&HERG&3.18\\
3C 330&16 09 35.01&+65 56 37.7&0.549&6.41&27.43&CD&HERG&2.81\\
3C 427.1&21 04 07.07&+76 33 10.8&0.572&6.54&27.53&CD&LERG&10.90\\
3C 295&14 11 20.65&+52 12 09.0&0.462&5.85&27.68&CD&LERG?&1.32\\
\enddata
\tablenotetext{a}{Jet, Fat Double, Classical Double}
\tablenotetext{b}{Value obtained by Konar et al. (2009)}
\label{tab:rawdata}
\end{deluxetable*}


\section{OBSERVATIONS AND DATA PREPARATION}

\subsection{X-ray data}

The X-ray observations for this study came from {\it Chandra} and {\it XMM-Newton}. The 3C sources had already been observed; the 6C, 7C and TOOT observations were made for this program. The {\it XMM-Newton} observations used the three EPIC cameras with the medium filter and the {\it Chandra} observations used the ACIS-S3 chip in either FAINT or VFAINT mode. Observation IDs and times are given in Table~\ref{tab:obsdata}.

We used the \textit{Chandra} analysis package \textsc{ciao} v4.3 for processing the \textit{Chandra} events file. We reprocessed the files using the \textit{chandra\_repro} script, applying particle background cleaning for observations in the VFAINT mode, and then checked for background flares by extracting a light curve using \textit{dmextract}. We excluded events at more than 3$\sigma$ above the background rate -- we removed a large flare from 3C~295 and small flares from 3C~16, 3C~200, 7C~0219+3423 and TOOT~1626+4523. Screened observation times are included in Table~\ref{tab:obsdata}.

The \textit{XMM-Newton} events files were reprocessed with the latest calibration data using \textit{XMM-Newton} \textsc{sas} v11.0.0. The pn camera data were filtered to include only single and double events (PATTERN $\leq$ 4) and data from the MOS cameras were filtered using the standard pattern mask (PATTERN $\leq$ 12). The data sets were also filtered to remove bad pixels, bad columns etc. We checked each events file for flares using the light curves at higher energy levels than those emitted by the sources (10--12 keV for the MOS cameras, 12--14 keV for the pn camera) and used good-time-interval (GTI) filtering to select data where the light curve was within 20\% of the quiescent level. We then used \textit{evigweight} to correct the events files for vignetting.

The particle background in the \textit{XMM-Newton} sources was removed using the method described by Croston et al. (2008a). This uses closed filter files (supplied courtesy of E. Pointecouteau) that were processed, filtered and weighted in the same manner as the source data sets. The closed filter data were rotated to match the source observations and scaling factors were calculated by comparing the count rates at 12--14 keV (pn camera) and 10--12 keV (MOS cameras). The closed filter data were then scaled by these factors before carrying out background subtraction when generating profiles and spectra.

For both the \textit{Chandra} and \textit{XMM-Newton} sources, we examined images of the sources in \textit{ds9}, overlaying them with the radio contours to find the approximate center of the extended source. We identified point sources in the data sets and emission associated with the radio lobes, which we excluded during subsequent analysis.


\renewcommand{\arraystretch}{1.0}
\begin{deluxetable*}{lcccc}\tabletypesize{\footnotesize}
\tablecolumns{5}
\tablewidth{0pc}
\tablecaption{Observation data for the ERA sample}
\tablehead{
	\colhead{Source}&
	\colhead{Instrument\tablenotemark{a}}&
	\colhead{Observation}&
	\colhead{Exposure\tablenotemark{b}}&
	\colhead{Screened\tablenotemark{b}}\\
	\colhead{}&
	\colhead{}&
	\colhead{}&
	\colhead{time (ks)}&
	\colhead{time (ks)}\\}

\startdata

TOOT 1301+3658&C&11568&39.5&39.5\\
TOOT 1255+3556&C&11569&39.5&39.5\\
TOOT 1626+4523&C&11570&36.7&36.4\\
TOOT 1630+4534&C&11571&20.8&20.8\\
TOOT 1307+3639 &C&11572&39.5&39.5\\
7C 0223+3415&X&551630101&41.0/41.0/30.6&32.5/35.7/21.4\\
7C 1731+6638&C&11573&24.0&24.0\\
7C 0213+3418&X&0551630201&43.9/44.0/33.9&15.4/27.1/3.4\\
TOOT 1303+3334&C&11574&39.7&39.7\\
7C 0219+3423&C&11575&39.3&39.3\\
6C 0850+3747&C&11576&39.2&39.2\\
6C 1200+3416&X&0551630301&49.6/49.7/39.0&37.6/41.0/24.6\\
6C 1132+3439&C&11577&39.6&39.6\\
6C 0857+3945&X&551630601&24.9/24.8/18.2&10.0/18.3/12.4\\
3C 16&C&13879&11.9&11.9\\
3C 46&X&0600450501&17.9/17.9/13.2&5.5/6.8/3.4\\
3C 341&X&0600450601&16.7/16.7/13.3&13.3/16.1/11.2\\
3C 200&C&838&14.7&14.7\\
3C 19&C&13880&11.9&11.9\\
3C 457&X&0502500101&52.2/52.2/40.0&36.8/39.2/28.8\\
3C 274.1&X&0671640801&27.2/27.1/22.6&22.5/22.9/16.6\\
3C 244.1&C&13882&11.9&11.8\\
3C 228&C&2453/2095&10.6/13.8&24.4\\
3C 330&C&2127&44.2&44.0\\
3C 427.1&C&2194&39.5&39.5\\
3C 295&C&578&18.8&18.2\\
\enddata
\tablenotetext{a}{\textit{Chandra}, \textit{XMM-Newton}}
\tablenotetext{b}{MOS1/MOS2/pn cameras for {\it XMM-Newton}}
\label{tab:obsdata}
\end{deluxetable*}


\subsection{Radio data}

Radio maps were used to make the overlay images shown in Appendix~\ref{app:sb} and to mask out the radio lobes so that any radio-related X-ray emission did not contaminate our measurements of the cluster properties. In many cases we used existing maps, either from the 3CRR Atlas\footnote{http://www.jb.man.ac.uk/atlas}; the Faint Images of the Radio Sky at Twenty Centimeters Survey (FIRST) (Becker
et al. 1995); Hardcastle et al. (2002b); and Croston et al. (2005b). For some of the less luminous sources where adequate maps were not available from these sources, we used the 1.4-GHz observations of Mitchell (2005) which were taken in A and C configurations. We obtained these from the VLA archive (Program ID AR477) and reduced them in {\sc aips} in the standard manner. Table~\ref{radiotable} contains full details of the radio maps used.


\renewcommand{\arraystretch}{1.0}
\begin{deluxetable}{llrrrrr}
\tablewidth{0pc} \tablecaption{Details of radio maps used in the analysis}
\tablehead{Source&Map frequency (GHz)&Resolution($\prime\prime$)&Ref.\\}
\startdata
TOOT\,1301$+$3658&$1.4$&$5.4 \times 5.4$&1\\
TOOT\,1255$+$3556&$1.4$&$5.4 \times 5.4$&1\\
TOOT\,1626$+$4523&$1.4$&$17 \times 13$&2\\
TOOT\,1630$+$4534&$1.4$&$1.3 \times 1.3$&2\\
TOOT\,1307$+$3639&$1.4$&$1.3 \times 1.3$&2\\
7C\,0223$+$3415&$1.4$&$14.7 \times 12.7$&2\\
7C\,1731$+$6638&$1.4$&$1.5 \times 1.2$&2\\
7C\,0213$+$3418&$1.4$&$14.6 \times 12.7$&2\\
TOOT\,1303$+$3334&$1.4$&$5.4 \times 5.4$&1\\
7C\,0219$+$3423&$1.4$&$1.4 \times 1.3$&2\\
6C\,0850$+$3747&$1.4$&$1.4 \times 1.3$&2\\
6C\,1200$+$3416&$1.4$&$5.4 \times 5.4$&1\\
6C\,1132$+$3439&$1.4$&$5.4 \times 5.4$&1\\
6C\,0857$+$3945&$1.4$&$5.4 \times 5.4$&1\\
3C\,16&1.4&$1.2 \times 1.2$&3\\
3C\,46&1.5&$4.2 \times 4.2$&3\\
3C\,341&1.4&$1.3 \times 1.3$&3\\
3C\,200&4.9&$0.33 \times 0.33$&3\\
3C\,19&1.5&$0.15 \times 0.15$&3\\
3C\,457&1.4&$5.1 \times 5.1$&3\\
3C\,274.1&$1.4$&$5.4 \times 5.4$&1\\
3C\,244.1&8.4&$0.75 \times 0.75$&4\\
3C\,228&8.4&$1.2 \times 1.2$&4\\
3C\,330&1.5&$1.5 \times 1.5$&5\\
3C\,427.1&1.5&$1.8 \times 1.1$&6\\
3C\,295&8.7&$0.2 \times 0.2$&3\\
\enddata
\label{radiotable}
\tablecomments{References: (1) Becker et al. (1995), (2) Mitchell
(2005), (3) \url{http://www.jb.man.ac.uk/atlas}, (4) Mullin et al. (2008), (5) Hardcastle et al. (2002b), (6) Croston et al. (2005b)}
\end{deluxetable}


\section{ANALYSIS}

The aim of the analysis was to find the temperature and X-ray luminosity of the ICM emission of the radio galaxies. Where possible, the temperature was obtained by spectral analysis; when there were insufficient counts, it was estimated from the count rate. The luminosity was determined by integrating the surface brightness profiles to the $R_{500}$ radius (defined in Section~\ref{sec:spat}).

\subsection{Imaging}

Appendix~\ref{app:sb} contains images of the X-ray emission of each cluster in the 0.5--5 keV energy range, with the radio emission overlaid as contours. We generated images for the {\it XMM-Newton} sources using the method described in Croston et al. (2008a). An image was extracted for each of the three EPIC cameras using {\it evselect}. The MOS images were then scaled to make their sensitivity equivalent to the pn camera image so that there would be no chip-gap artifacts when the three images were combined. We generated exposure maps for each camera using {\it eexpmap}, which were used to correct for the chip gaps, but not for vignetting as this leads to incorrect scaling of the particle background that dominates at large radii. The resulting images are therefore not vignetting corrected; they are purely pictorial and not used in any subsequent analysis.

For both {\it Chandra} and {\it XMM-Newton} sources, we used {\it dmfilth} to replace point sources by Poission noise at the level of nearby regions, and then applied Gaussian smoothing using {\it aconvolve}. We then used {\it ds9} to display the X-ray emission and overlay the radio emission contours.

\subsection{Spatial analysis}\label{sec:spat}
We extracted a radial surface brightness profile from the events file of each source by taking the average counts in annuli around the source centroid. We used an annulus outside the maximum detection radius to obtain the background counts, and these were subtracted from the source counts to obtain the net counts in each annulus. The point sources and radio emission identified during data preparation were removed and the annulus areas adjusted to account for the excluded regions. We used an energy range of 0.4--7.0 keV, this being within the reliable operating range for the \textit{Chandra} data. For one source (3C~341), the extended emission was swamped by a bright nucleus, and we reduced the energy range to 0.4--2.0 keV to cut down the nuclear emission so that the profile could be modeled.

For the \textit{XMM-Newton} sources, since the pn camera is more sensitive than the MOS cameras, we created the pn profile first and then used the same annuli and background area for the MOS profiles. The three profiles were then scaled by their relative exposure times and added together.

Table~\ref{tab:beta} contains the maximum detection radius and net counts within that radius for each of the sources.


\renewcommand{\arraystretch}{1.0}
\begin{deluxetable*}{lccccrr}\tabletypesize{\footnotesize}
\tablecolumns{7}
\tablewidth{0pc}
\tablecaption{Radial profile modeling results}
\tablehead{
	\colhead{Source}&
	\colhead{R$_{det}$}&
	\colhead{Counts}\tablenotemark{a}&
	\colhead{PSF fit}&
	\colhead{PSF+$\beta$ fit}&
	\colhead{$\beta$}\tablenotemark{b}&
	\colhead{r$_c$}\tablenotemark{b}\\
	\colhead{}&
	\colhead{kpc}&
	\colhead{}&
	\colhead{$\chi^2$/dof}&
	\colhead{$\chi^2$/dof}&
	\colhead{}&
	\colhead{arcsec}\\}
\startdata
TOOT 1301+3658&&${\it (<94)}$&\multicolumn{2}{c}{{\it Low counts}}&{\it 0.49}&{\it 5.58} \\
TOOT 1255+3556&174&340&5.42/5&0.23/5&$1.48^{+0.02}_{-1.15}$&$19.00^{+11.6}_{-16.5}$ \\
TOOT 1626+4523&286&108&15.9/6&0.617/6&$0.30^{+0.19}_{-0.10}$&$2.10^{+7.89}_{-1.10}$ \\
TOOT 1630+4534&&${\it (<63)}$&\multicolumn{2}{c}{{\it Low counts}}&{\it 0.49}&{\it 5.13} \\
TOOT 1307+3639&162&26&4.18/4&0.007/4&$0.53^{+1.08}_{-0.16}$&$1.91^{+7.89}_{-0.90}$ \\
7C 0223+3415&&${\it (<77)}$&\multicolumn{2}{c}{{\it Low counts}}&{\it 0.49}&{\it 5.24} \\
7C 1731+6638&&${\it (<63)}$&\multicolumn{2}{c}{{\it Low counts}}&{\it 0.49}&{\it 4.79} \\
7C 0213+3418&&${\it (<100)}$&\multicolumn{2}{c}{{\it Low counts}}&{\it 0.49}&{\it 5.29} \\
TOOT 1303+3334&68.9&867&32.8/7&6.14/7&$0.22^{+0.13}_{-0.05}$&$7.82^{+2.17}_{-6.55}$ \\
7C 0219+3423&164&46&6.0/4&0.008/4&$0.45^{+0.54}_{-0.24}$&$4.70^{+5.28}_{-3.70}$ \\
6C 0850+3747&534&2351&84.0/9&10.2/9&$0.45^{+0.04}_{-0.02}$&$1.01^{+1.49}_{-0.01}$ \\
6C 1200+3416&1007&1983&120/10&36.5/10&$0.41^{+0.03}_{-0.04}$&$6.22^{+0.07}_{-4.65}$ \\
6C 1132+3439&669&389&55.2/10&9.30/10&$0.38^{+0.07}_{-0.09}$&$7.73^{+2.26}_{-6.73}$ \\
6C 0857+3747&816&612&8.75/9&2.17/9&$0.41^{+0.15}_{-0.17}$&$1.12^{+15.80}_{-0.12}$ \\
3C 16&&${\it (<60)}$&\multicolumn{2}{c}{{\it Low counts}}&{\it 0.49}&{\it 5.74} \\
3C 46&510&170&13.4/5&0.57/5&$1.49^{+0.01}_{-0.97}$&$66.3^{+18.0}_{-45.5}$ \\
3C 341&517&214&20.2/5&0.25/5&$0.54^{+0.96}_{-0.10}$&$1.74^{+40.36}_{-0.23}$ \\
3C 200&200&259&16.1/6&0.77/6&$0.48^{+0.15}_{-0.05}$&$1.10^{+1.78}_{-0.09}$ \\
3C 19&589&503&294/8&4.38/8&$0.61^{+0.06}_{-0.05}$&$7.73^{+1.91}_{-1.67}$ \\
3C 457&1119&2402&51.4/9&6.46/9&$0.45^{+0.21}_{-0.11}$&$18.3^{+29.6}_{-17.3}$ \\
3C 274.1&610&678&121/8&3.80/8&$1.16^{+0.34}_{-0.41}$&$32.0^{+11.5}_{-13.4}$ \\
3C 244.1&966&171&27.8/6&0.818/6&$0.41^{+0.04}_{-0.05}$&$1.03^{+2.13}_{-0.01}$ \\
3C 228&537&768&32.5/9&3.28/9&$0.90^{+1.10}_{-0.33}$&$12.3^{+15.6}_{-6.2}$ \\
3C 330&473&360&82.1/10&1.28/10&$0.56^{+0.11}_{-0.07}$&$4.80^{+2.65}_{-1.86}$ \\
3C 427.1&675&721&288/10&4.72/10&$0.40^{+0.14}_{-0.02}$&$3.14^{+1.22}_{-1.18}$ \\
3C 295&719&5308&3978/17&37.37/17&$0.50^{+0.005}_{-0.004}$&$2.52^{+0.19}_{-0.15}$ \\
\enddata
\tablenotetext{a}{Upper limits were obtained within estimated R$_{500}$. Counts for {\it XMM-Newton} sources are for the pn camera only}
\tablenotetext{b}{Italics indicate median values used for sources with low counts.}
\label{tab:beta}
\end{deluxetable*}

We fitted the surface brightness profiles with $\beta$ models (see below) using the Markov-Chain Monte Carlo (MCMC) method described by Croston et al. (2008a) to explore the parameter space of these models and thus find Bayesian estimates of the core radii ($r_{c}$) and $\beta$ values. This MCMC method uses the Metropolis-Hastings algorithm in a manner similar to the METRO code by Hobson \& Baldwin (2004), but implemented to run on a cluster of multiple processors using the Message Passing Interface (MPI: See Mullin \& Hardcastle (2009) for more details). The method we use here differs from the the implementation of this method used by Croston et al. (2008a), in that we use a new fitting engine which allows the normalization of the $\beta$ model and of any point-source component to vary freely during the fits. Plausible ranges for the free parameters were estimated and used to define uniform (uninformative) priors for the MCMC method: for the normalizations and core radii (see below), priors uniform in log space were used to avoid bias towards large values. The uncertainties corresponding to 1$\sigma$ errors for 2 interesting parameters were determined using 1-dimensional projections of the minimal n-dimensional volume that encloses 68\% of the posterior probability distribution as returned by the MCMC code. This code was also used to determine the luminosities as discussed below.

We initially used the appropriate instrument point spread function (PSF) alone to check whether this gave a satisfactory fit to the data. We then added a single $\beta$ model to fit the extended emission from the ICM (Cavaliere \& Fusco-Femiano 1976), convolved with the PSF. The surface brightness at radius $R$ is given by

$S(R)=S_0(1+(\frac{R}{r_{\rm c}})^2)^{-3\beta+0.5}$

\noindent where $r_c$ is the core radius. Although the surface brightness distributions obtained from high-resolution / high sensitivity observations of many clusters have been found to be more complex than a single-fit $\beta$-model, such a model provides an adequate parametrization of the surface brightness profile to integrate to obtain luminosity, given the comparatively poor data quality at these redshifts. The goodness-of-fit and $\beta$ model parameters are shown in Table~\ref{tab:beta} and Appendix~\ref{app:sb} contains the surface brightness profiles overlaid with the PSF and $\beta$ model profiles. 

We could not obtain fits for six sources. TOOT~1630+4534 could not be detected above the background (neither nuclear or extended emission); TOOT~1301+3658, 7C~0213+3418 and 3C~16 had insufficient counts to create a profile, and 7C~0223+3415 and 7C 1731+6638 had a point-source detection, but no extended emission above the level of the PSF wings. For these sources we estimated a 3$\sigma$ upper limit on the counts by obtaining the net counts within an estimated R$_{500}$ overdensity radius (see below). For the four faint sources, we used the median R$_{500}$ of the 7C and TOOT sources; for 3C~16 we used the median of the 6C and 3C sources. If the net counts were greater than three times the error on the counts in the background area, we used the net counts plus three times the error for the upper limit on the counts; otherwise we used three times the background error. (Since 7C 1731+6638 has a strong psf with no detectable extended emission, we used the background error method rather than the net counts method). 

\begin{figure*}
\plottwo{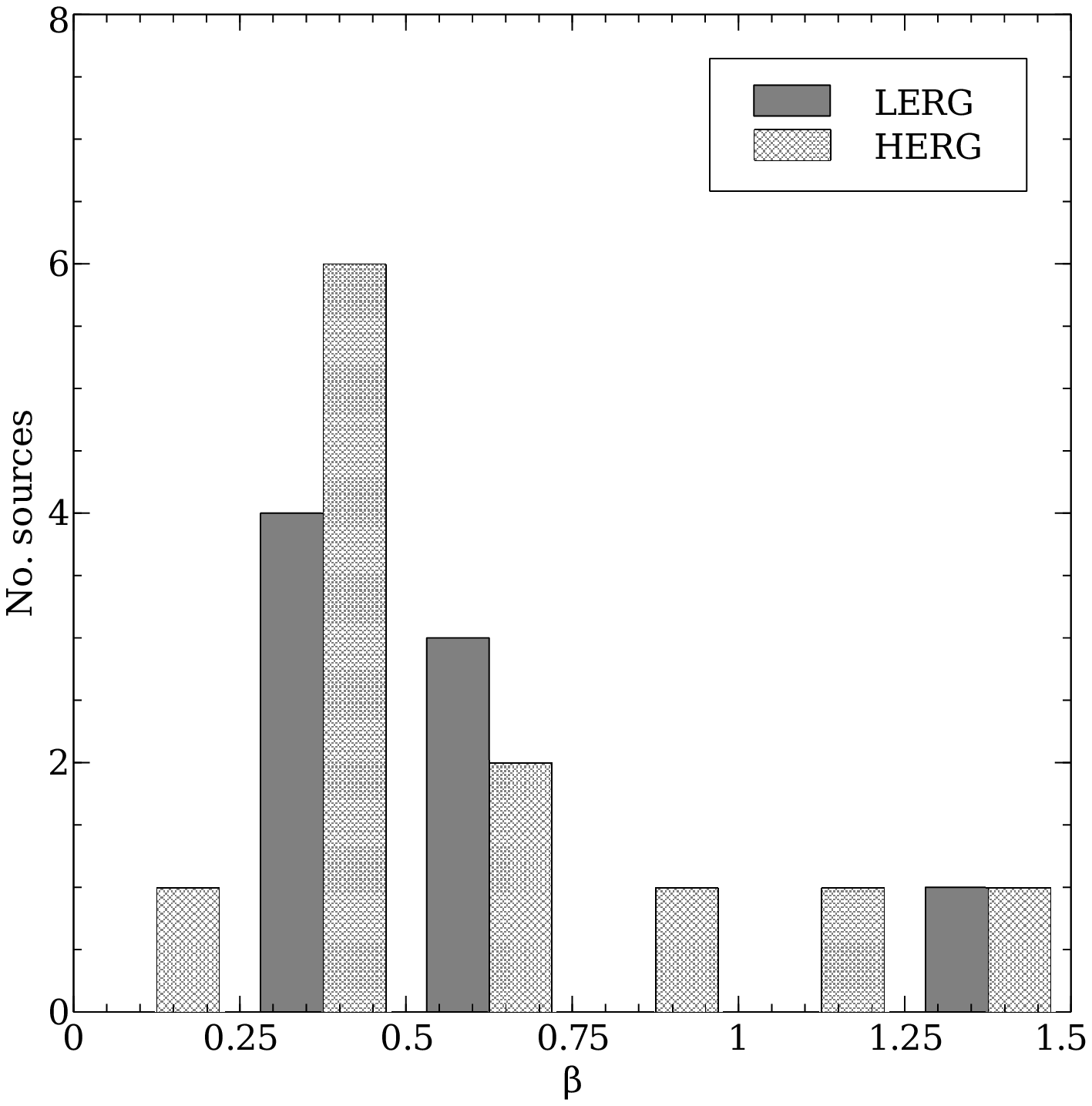}{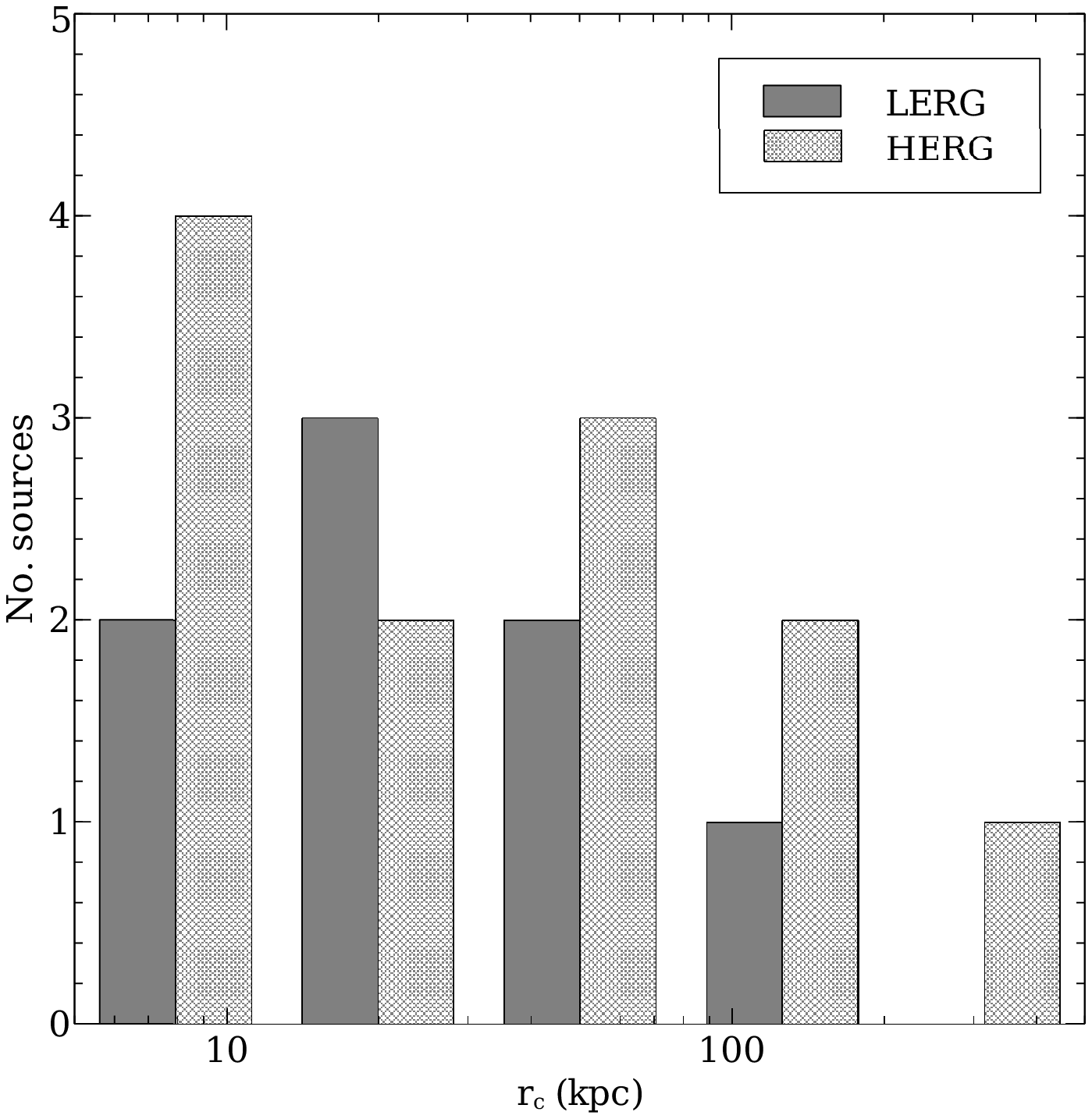}
\caption{Distribution of the $\beta$ model parameters; $\beta$ (left) and core radius (right).}
\label{fig:betas}\end{figure*}

The distributions of $\beta$ and the core radius, $r_c$, are shown in Figure~\ref{fig:betas}. The majority of the $\beta$ values are around 0.5, which is expected for groups and poor clusters (eg Helsdon \& Ponman 2000), but there are three very high values of $\beta$ (TOOT~1255+3556, 3C~46 and 3C~274.1 -- these also have the three highest core radii) and two very low values (TOOT~1626+4523 and TOOT~1303+3334). The three TOOT sources are faint objects with low counts, so the model parameters are very poorly constrained, and McLure et al. (2004) identified the host galaxy of 3C~46 as having undergone a major merger so its ICM may not be in hydrostatic equilibrium. 3C~274.1, however, has an undisturbed host galaxy and so its steep profile is perhaps surprising. However, the profile fits the data well and since we are using the profile simply to obtain luminosity we are concerned only with the shape of the profile and not the accuracy of the $\beta$ model. The $\beta$ and $r_c$ values are degenerate and therefore not physically very meaningful, and the uncertainties on the extreme values are large and so cover more typical values.

Luminosity was calculated by integrating the $\beta$ model profile to the $R_{500}$ overdensity radius, using counts to flux conversion factors generated from the {\it apec} model. We extrapolated the $\beta$ model to the $R_{500}$ radius calculated using the $R-T$ relationship from Arnaud et al. (2005):

$R_{500}=h(z)^{-1}B_{500}(\frac{kT}{5})^{\beta}$, where $h^2(z)=\Omega_{\rm m}(1+z)^3+\Omega_\Lambda$

We calculated a luminosity for each sample of the output of the MCMC code, which provided us with a posterior probability distribution function over luminosity, marginalized over all other parameters. We used the median rather than the mean of the posterior probability
distribution function as our luminosity estimate because the distributions were skewed for the fainter sources. Our quoted uncertainties on the luminosity are credible intervals defined on this one-dimensional posterior probability distribution function such that
68\% of the probability is contained in the smallest luminosity range. The luminosity uncertainties take into account the (in some cases large) uncertainties on $\beta$ and $r_c$.

Table~\ref{tab:lum} contains the X-ray luminosities for each source: within the 0.4 to 7.0 keV energy range to the maximum detection radius; bolometric luminosity within the maximum detection radius; bolometric luminosity within $R_{500}$; and the bolometric luminosity scaled by $h^{-1}(z)$ to correct for the critical density evolution.

Nine sources have $R_{500}$ greater than the maximum detection radius. For these, the extrapolated counts for the luminosities within $R_{500}$ are less than 27\% greater than the observed counts for all but one source -- TOOT~1626+4523 -- which has a shallow surface brightness profile and $R_{500}$ much larger than the maximum detection radius. Consequently the $R_{500}$ luminosity for that source is more than twice the luminosity within the maximum detection radius. The mean extrapolation for the remaining eight sources is 12\%.


\renewcommand{\arraystretch}{1.0}
\begin{deluxetable*}{lllcll}\tabletypesize{\footnotesize}
\tablecolumns{6}
\tablewidth{0pc}
\tablecaption{ICM X-ray luminosity}
\tablehead{
	\colhead{Source}&
	\colhead{$L_X (0.4-7.0 keV)$}&
	\colhead{$L_X (bol)$}&
	\colhead{$R_{500}$}&
	\colhead{$L_X (bol)$}&
	\colhead{$h(z)^{-1}L_X (bol)$}\\
	\colhead{}&
	\colhead{$D_{rad}$}&
	\colhead{$D_{rad}$}&
	\colhead{$kpc$}&
	\colhead{$R_{500}$}&
	\colhead{$R_{500}$}\\
	\colhead{}&
	\colhead{$\times{10}^{43} erg.s^{-1}$}&
	\colhead{$\times{10}^{43} erg.s^{-1}$}&
	\colhead{}&
	\colhead{$\times{10}^{43} erg.s^{-1}$}&
	\colhead{$\times{10}^{43} erg.s^{-1}$}\\}

\startdata
TOOT 1301+3658&&&380\tablenotemark{a}&$<1.26$&$<1.01$\\
TOOT 1255+3556&$0.25^{+0.10}_{-0.24}$&$0.44^{+0.18}_{-0.42}$&351&$0.48^{+0.20}_{-0.47}$&0.37\\
TOOT 1626+4523&$1.61^{+0.70}_{-0.99}$&$2.45^{+1.07}_{-1.52}$&533&$5.30^{+2.83}_{-4.26}$&4.15\\
TOOT 1630+4534&&&380\tablenotemark{a}&$<2.44$&$<1.87$\\
TOOT 1307+3639&$0.23^{+0.11}_{-0.22}$&$0.40^{+0.20}_{-0.39}$&317&$0.41^{+0.20}_{-0.41}$&0.30\\
7C 0223+3415&&&380\tablenotemark{a}&$<1.45$&$<1.12$\\
7C 1731+6638&&&380\tablenotemark{a}&$<2.97$&$<2.19$\\
7C 0213+3418&&&380\tablenotemark{a}&$<8.51$&$<6.60$\\
TOOT 1303+3334&$3.92^{+1.89}_{-2.48}$&$6.00^{+2.90}_{-3.80}$&525&$7.54^{+3.98}_{-4.97}$&5.55\\
7C 0219+3423&$0.56^{+0.24}_{-0.53}$&$1.00^{+0.42}_{-0.95}$&380&$1.27^{+0.71}_{-1.25}$&0.92\\
6C 0850+3747&$2.95^{+0.42}_{-0.49}$&$4.22^{+0.60}_{-0.70}$&648&$4.53^{+0.70}_{-0.77}$&3.65\\
6C 1200+3416&$8.48^{+1.05}_{-1.04}$&$12.69^{+1.57}_{-1.55}$&553&$8.28^{+0.95}_{-0.92}$&6.21\\
6C 1132+3439&$6.61^{+1.81}_{-2.00}$&$10.89^{+2.98}_{-3.30}$&454&$7.30^{+1.61}_{-1.84}$&5.54\\
6C 0857+3945&$1.89^{+1.00}_{-1.42}$&$3.15^{+1.67}_{-2.35}$&447&$2.87^{+1.62}_{-2.06}$&2.16\\
3C 16&&&515\tablenotemark{b}&$<2.79$&$<2.25$\\
3C 46&$3.90^{+0.87}_{-1.09}$&$5.95^{+1.33}_{-1.67}$&535&$6.03^{+1.37}_{-1.72}$&4.79\\
3C 341&$0.79^{+0.34}_{-0.32}$&$1.33^{+0.57}_{-0.54}$&429&$1.29^{+0.51}_{-0.54}$&1.02\\
3C 200&$1.41^{+0.42}_{-0.41}$&$2.28^{+0.69}_{-0.67}$&474&$2.64^{+0.93}_{-0.99}$&2.07\\
3C 19&$24.643^{+1.90}_{-2.18}$&$36.589^{+2.82}_{-3.23}$&570&$36.4^{+2.8}_{-3.1}$&28.1\\
3C 457&$6.48^{+1.67}_{-1.48}$&$9.03^{+2.33}_{-2.06}$&670&$6.14^{+1.05}_{-1.13}$&4.89\\
3C 274.1&$2.25^{+0.22}_{-0.21}$&$3.82^{+0.38}_{-0.36}$&343&$3.66^{+0.35}_{-0.29}$&2.92\\
3C 244.1&$4.06^{+1.40}_{-1.55}$&$6.15^{+2.11}_{-2.35}$&529&$4.54^{+1.36}_{-1.33}$&3.61\\
3C 228&$2.09^{+0.44}_{-0.58}$&$3.22^{+0.67}_{-0.89}$&515&$3.22^{+0.67}_{-0.89}$&2.39\\
3C 330&$2.75^{+0.42}_{-0.55}$&$4.75^{+0.72}_{-0.95}$&430&$4.64^{+0.70}_{-0.86}$&3.44\\
3C 427.1&$19.5^{+1.8}_{-2.2}$&$27.9^{+2.5}_{-3.1}$&620&$26.2^{+2.5}_{-2.5}$&19.2\\
3C 295&$125^{+2}_{-2}$&$171^{+3}_{-3}$&966&$183^{+4}_{-4}$&143\\

\enddata
\tablenotetext{a}{Median $R_{500}$ of 7C and TOOT sources.}
\tablenotetext{b}{Median $R_{500}$ of 6C and 3C sources.}
\label{tab:lum}
\end{deluxetable*}


\subsection{Spectral analysis}

Where possible, we used spectral analysis to obtain the ICM temperature, using the \textit{apec} model for the thermal bremsstrahlung from the ICM and the \textit{wabs} photo-electric absorption model to take account of Galactic absorption. The high energy emission from the nucleus was excluded by reducing the energy range to 0.5 to 2.5 keV for all sources (0.5 to 2.0 keV for 3C~341 -- see Section~\ref{sec:spat}). In several of the \textit{XMM-Newton} sources, a 1.5 keV instrumental aluminium fluorescence line was visible, so we also excluded the 1.4 to 1.6 keV energy band for all the \textit{XMM-Newton} spectra (Freyberg et al. 2002).

The spectrum was extracted from an annulus excluding the central nucleus and extending to the maximum detection radius, using the same background annulus as the surface brightness profile. We estimated the size of a region appropriate for excluding the nucleus from the profile model fit by looking for the point at which the extended emission begins to dominate over the PSF. Non-nuclear point sources and lobe-related emission were also excluded. We used a metallicity of 0.3~solar (Balestra et al. 2007) for all sources except for 3C~295, 3C~330 and 3C~457, where metallicities had been calculated elsewhere (Belsole et al. 2007; Konar et al. 2009). These were 0.48, 0.2 and 0.35 respectively. We discuss the effect of varying metallicity in Section~\ref{sec:LrLx}.

The ICM spectrum for the \textit{Chandra} data could be fit immediately. For the \textit{XMM-Newton} sources we first modeled the Galactic X-ray background in each camera using \textit{apec} models for the two thermal components and a photon power-law with Galactic absorption for the extragalactic background. These were scaled to reflect the contribution from each camera, and the source spectra for the three cameras were then fit in the same manner as the \textit{Chandra} sources, but with the three background components.

Once we had an initial estimate for the temperature, we varied the inner and outer radii to check that we had a stable value for the temperature with reasonable error limits, suggesting that the isothermal assumption was applicable across that annulus. For example, 3C~295 is known to have a cool core -- Allen et al. (2001) found that the temperature rises with radius until about 10 arcsec.

There were sufficient counts to model the spectrum for ten of the sources. For the fainter sources with a $\beta$ model fit, we extracted a spectrum and used the count rate from the $\beta$ model to estimate luminosity from the \textit{apec} model. This was converted to temperature using the scaling relation of Pratt et al. (2009), which provided an initial estimate of R$_{500}$ (see Section~\ref{sec:spat}). Iterating the process gave new estimates of the counts within R$_{500}$, luminosity and temperature until the temperature converged. For the {\it XMM-Newton} sources, we used counts from the pn camera only in this process. The same process was used for the sources without a $\beta$ model fit, using the 3$\sigma$ upper limit on the counts.

Table~\ref{tab:temp} contains the inner and outer radii of the annuli, the temperatures of the sources and the $\chi^2$ for the temperatures obtained by spectral analysis. As expected, the temperatures are for the most part typical of poor clusters with only three sources above 3 keV. The errors on most of the spectral temperatures are large, as is to be expected from such faint objects. Temperatures have been obtained for some of the clusters by other researchers (Belsole et al. 2007; Allen et al. 2001; Konar et al. 2009) and our values are compatible with their results.

We have not included any corrections for the reported difference between temperatures obtained by {\it Chandra} and {\it XMM-Newton} since the difference at 3 keV and less is slight (Schellenberger et al. 2012). 


\renewcommand{\arraystretch}{1.0}
\begin{deluxetable}{llccrc}\tabletypesize{\footnotesize}
\tablecolumns{6}
\tablewidth{0pc}
\tablecaption{ICM temperatures}
\tablehead{
	\colhead{Source}&
	\colhead{Method}&
	\colhead{Annulus radii}&
	\colhead{Temperature}&
	\colhead{$\chi^2$/dof}\tablenotemark{a}\\
	\colhead{}&
	\colhead{}&
	\colhead{kpc}&
	\colhead{keV}&
	\colhead{}\\}

\startdata

TOOT 1301+3658&Upper limt&&$<1.40$& \\
TOOT 1255+3556&Estimate&15--174&$1.04^{+0.11}_{-0.63}$& \\
TOOT 1626+4523&Estimate&15--296&$2.14^{+0.29}_{-0.81}$& \\
TOOT 1630+4534&Upper limt&&$<1.69$& \\
TOOT 1307+3639&Estimate&16--162&$0.98^{+0.11}_{-0.67}$& \\
7C 0223+3415&Upper limit&&$<1.41$& \\
7C 1731+6638&Upper limit&&$<1.77$& \\
7C 0213+3418&Upper limit&&$<2.37$& \\
TOOT 1303+3334&Estimate&51--448&$2.33^{+0.32}_{-0.62}$& \\
7C 0219+3423&Estimate&17--164&$1.36^{+0.18}_{-0.91}$& \\
6C 0850+3747&Spectrum&21--160&$2.86^{+2.11}_{-0.78}$&3.77/4 \\
6C 1200+3416&Spectrum&252--944&$2.46^{+1.89}_{-0.65}$&84.0/78 \\
6C 1132+3439&Spectrum&15--486&$1.71^{+0.69}_{-0.37}$&4.16/5 \\
6C 0857&Estimate&251--816&$1.69^{+0.24}_{-0.55}$& \\
3C 16&Upper limit&&$<1.78$& \\
3C 46&Estimate&170--510&$2.11^{+0.14}_{-0.20}$& \\
3C 341&Estimate&115-517&$1.45^{+0.16}_{-0.22}$& \\
3C 200&Estimate&15-215&$1.74^{+0.16}_{-0.23}$& \\
3C 19&Spectrum&29-383&$2.47^{+0.63}_{-0.38}$&4.78/11 \\
3C 457&Spectrum&168--839&$3.10^{+2.95}_{-1.00}$&95.7/75 \\
3C 274.1&Spectrum&166--555&$0.95^{+0.29}_{-0.25}$&9.99/12 \\
3C 244.1&Estimate&14--966&$2.05^{+0.17}_{-0.19}$& \\
3C 228&Spectrum&31--316&$2.22^{+2.39}_{-0.71}$&1.06/4 \\
3C 330&Spectrum&25--441&$1.61^{+1.26}_{-0.35}$&0.76/3 \\
3C 427.1&Spectrum&32--193&$3.14^{+5.27}_{-1.18}$&2.86/4 \\
3C 295&Spectrum&29--575&$6.09^{+0.73}_{-0.73}$&72.3/85 \\

\enddata
\tablenotetext{a}{For temperatures from spectral analysis}
\label{tab:temp}
\end{deluxetable}

\section{RESULTS AND DISCUSSION}

\subsection{Radio galaxy environments}\label{sec:LrLx}

\begin{figure*}
\plottwo{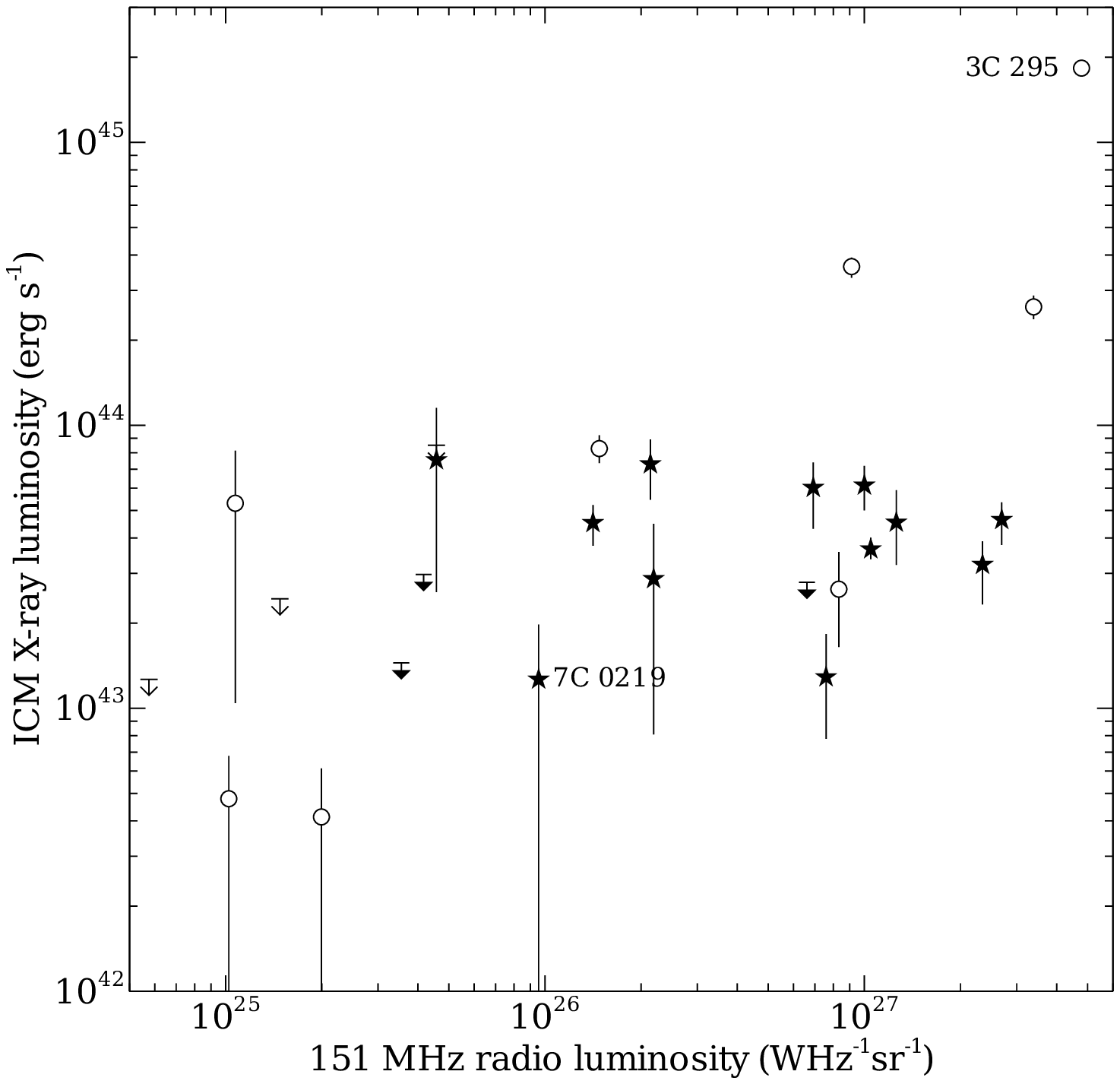}{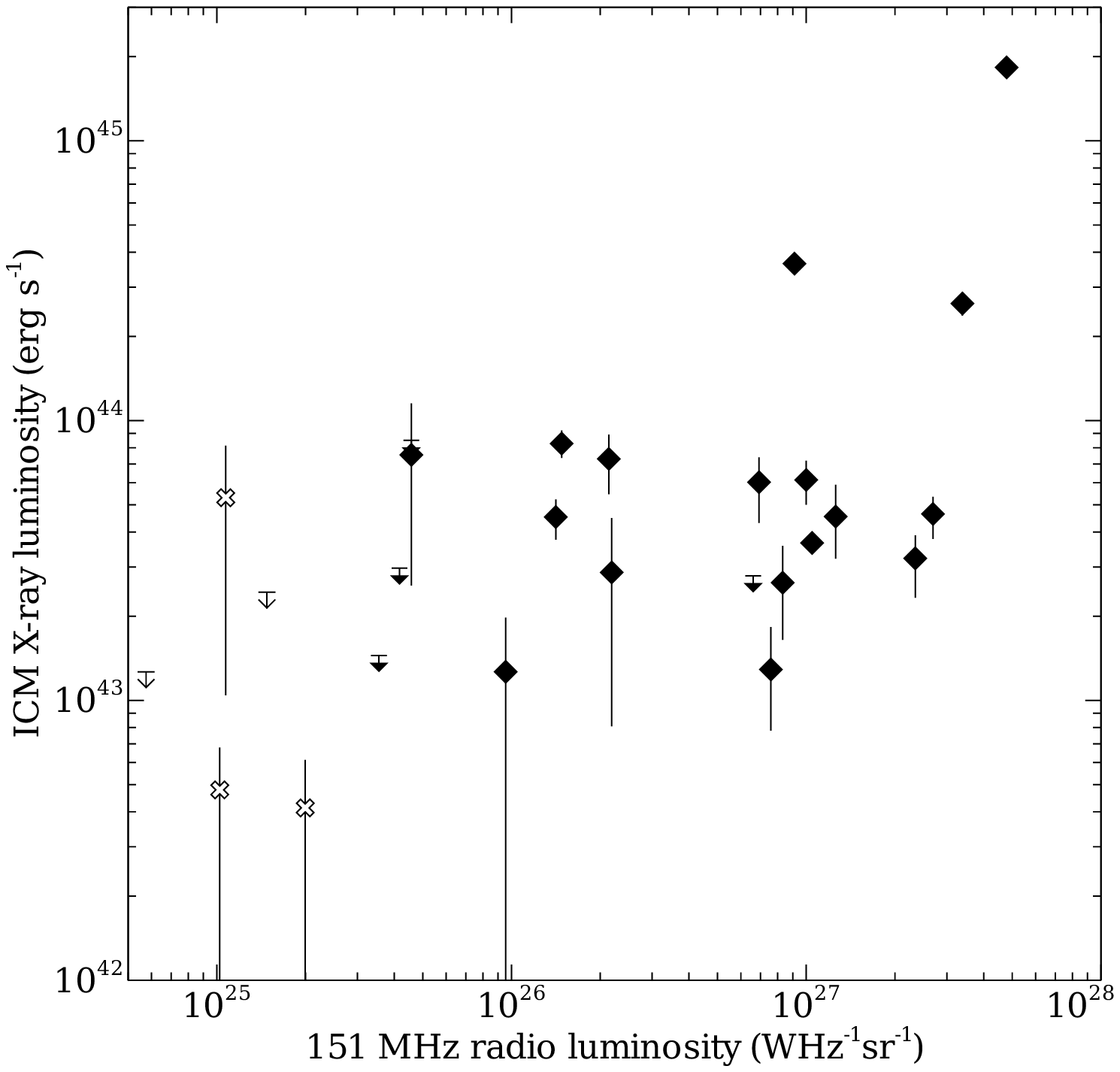}
\caption{Radio luminosity vs ICM X-ray luminosity, separated into excitation classes (left), and FRI and FRII galaxies (right). In the left plot, open circles are LERGs, open arrows are LERG upper limits, filled stars are HERGs and filled arrows are HERG upper limits. In the right plot, open crosses are FRIs, open arrows are FRI upper limits, filled diamonds are FRIIs and filled arrows are FRII upper limits.}
\label{fig:LxLrHL}\end{figure*}

\begin{figure}
\plotone{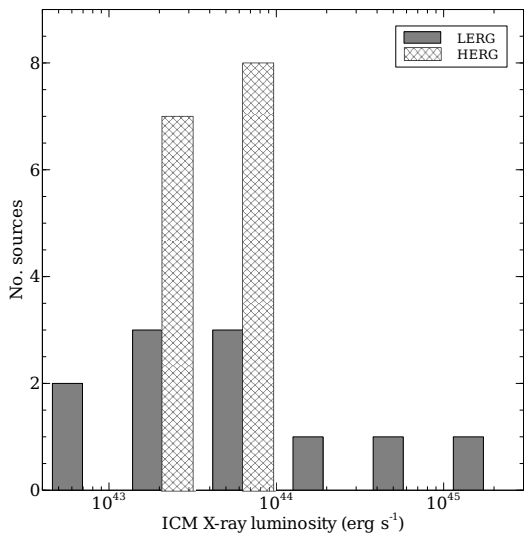}
\caption{ICM X-ray luminosity coverage, separated into excitation classes.}
\label{fig:Lxhisto}\end{figure}

In order to investigate the relationship between radio galaxies and their cluster environments, we compared radio luminosity and ICM luminosity for the full sample, for the different excitation classes and for the FR classes. These are plotted in Figures~\ref{fig:LxLrHL} and \ref{fig:Lxhisto}.

The majority of the sources have similar ICM luminosity. Although the radio luminosity covers three decades, 80\% of the sources have ICM luminosity within the single decade from $10^{43}$ to $10^{44}$ erg~s$^{-1}$. The five sources outside this range are all LERGs (with the possible exception of 3C~295), spreading over about 2.5 decades, and X-ray luminosity does appear to rise with radio luminosity for the LERGS. Since we had censored data, we checked for correlations using generalized Kendall's $\tau$ tests (Isobe et al. 1986), and the results are shown in Table~\ref{tab:stats}. We found a correlation between radio luminosity and ICM luminosity for the full sample and for the LERG subsample, but not for the HERGs. The LERG result does however depend heavily on 3C~295; the correlation is considerably weaker when 3C~295 is removed from the LERG subsample.

We checked that this result was independent of black hole mass using the black hole masses calculated by McLure et al. (2004), who found a correlation between radio luminosity and black hole mass that was driven by the HERG subsample. We found no evidence of a correlation between ICM luminosity and black hole mass for the full sample or for the individual subsamples.

Since the extrapolation of the luminosity to R$_{500}$ depends on temperature, we checked the effect of propagating the temperature uncertainties through the calculations for some of the sources with large temperature errors and/or big differences between the maximum detection radius and R$_{500}$. For most of the sources, the difference was slight, but there were big changes to the uncertainties for some of the sources with large errors on the temperatures. 3C~427.1 had the biggest change, with the 1$\sigma$ range going from $28-30\times10^{43}$ erg~s$^{-1}$ to $19-54\times10^{43}$ erg~s$^{-1}$. This would however retain 3C~427.1's position as one of the most luminous clusters and radio galaxies and so the LERG correlation remains. 3C~457 has the next largest temperature errors and the upper limit on the luminosity potentially increases from $7\times10^{43}$ to $11\times10^{43}$ erg~s$^{-1}$. Its position within Figure~\ref{fig:LxLrHL}, however, means that an increase in luminosity would merely weaken the HERG correlation further. Overall, we found that propagating the temperature uncertainties through the calculations would not affect the results of the correlation tests.

A second potential source of error was our assumption of a metallicity of 0.3~solar for the majority of the sources. There were insufficient counts to allow metallicity to vary during modeling, so we recalculated the temperatures and luminosities of three of our sources with spectral temperatures (3C~427.1, 6C~1132+3439 and 6C~1200+3416) with metallicities of 0.1 and 0.5~solar, these being outside the rms bounds given by Balestra et al. (2007). The biggest change in luminosity was 4\%, which was well within the luminosity errors of the sources. We also recalculated the luminosities for three of the sources with estimated temperatures. The biggest luminosity change was for the coolest source -- TO~1255+3556 -- where dropping the metallicity to 0.1~solar increased the luminosity by 12\%. Again, this was well within the 1$\sigma$ error bounds. We therefore concluded that inaccuracies in our metallicities were unlikely to affect our results.

As can be seen from the right-hand plot in Figure~\ref{fig:LxLrHL}, any correlation for the FRII subgroup would be weak, and this is confirmed by the Kendall's $\tau$ test. There is insufficient data to examine the FRI subgroup.


\renewcommand{\arraystretch}{1.0}
\begin{deluxetable}{lccc}\tabletypesize{\footnotesize}
\tablecolumns{4}
\tablewidth{0pc}
\tablecaption{Generalized Kendall's $\tau$ correlation tests}
\tablehead{
	\colhead{Subpopulation}&
	\colhead{Z}&
	\colhead{N}&
	\colhead{p}\\}

\startdata
\sidehead{Radio luminosity vs ICM luminosity}
All data&2.841&26&0.0045 \\
HERG&0.862&15&0.3887 \\
HERG--no 7C~0219&0.446&14&0.6557 \\
LERG&2.655&11&0.0079 \\
LERG--no 3C~295&2.140&10&0.0324 \\
FRII&1.903&21&0.0571 \\
\sidehead{Black hole mass vs ICM luminosity}
All data&1.539&26&0.1237 \\
HERG&1.470&15&0.1415 \\
LERG&1.202&11&0.2292 \\
\sidehead{ICM temperature vs luminosity}
All data&5.082&26&$<0.00005$ \\
Spectrum method&2.236&9&0.0253 \\
\sidehead{B$_{gg}$(977 kpc) vs ICM luminosity}
All data&1.965&26&0.0494 \\
HERG&0.963&15&0.3354 \\
LERG&1.456&11&0.1454 \\
FRII&2.064&21&0.0390 \\
\sidehead{B$_{gg}$(564 kpc) vs ICM luminosity}
All data&2.557&26&0.0106 \\
HERG&1.878&15&0.0604 \\
LERG&1.456&11&0.1454 \\
FRII&3.064&21&0.0022 \\

\enddata
\label{tab:stats}
\tablecomments{Z is the correlation statistic; N is sample size; p is probability under the null hypothesis}
\end{deluxetable}

\subsection{Comparison with general cluster environments}

During the analysis, we used various assumptions based on the expectation that the cluster environments of our radio-loud AGN are not markedly different from other clusters of similar luminosity. In particular we assumed that the X-ray cluster luminosity is a good proxy for gravitational mass and that the ICM follows the expected luminosity--temperature relation. There is evidence that radio-loud groups of a given X-ray luminosity are hotter than similar radio-quiet groups (Croston et al. 2005a), but this effect is small enough not to be seen at higher luminosities (Belsole et al. 2007).

\begin{figure}
\plotone{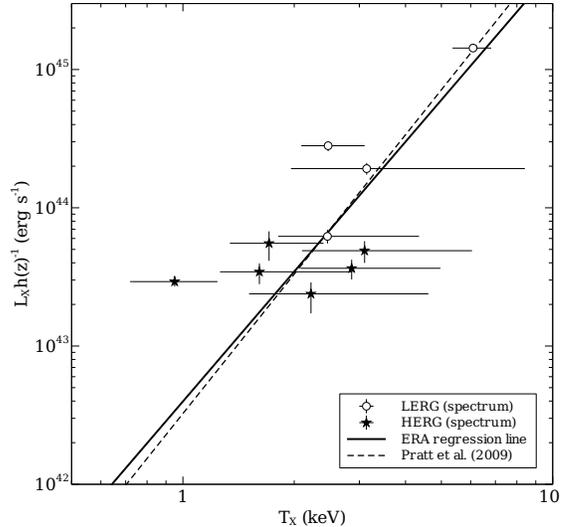}
\caption{ICM luminosity vs temperature for the temperatures obtained by spectral analysis. The heavy, solid line shows the $L_{\rm X}-T_{\rm X}$ relationship from all the ERA results (including estimated temperatures and upper limits) and the dashed line shows Pratt et al. (2009)'s $L_{\rm X}-T_{\rm X}$ relation.}
\label{fig:LxTx}\end{figure}

We expected that our cluster environments, although relatively poor, would be sufficiently luminous for any disruption by feedback to be smaller than the experimental errors. We checked the validity of this assumption by looking at the $L_{\rm X}-T_{\rm X}$ relation, the gravitational masses and the entropy of the clusters. Figure~\ref{fig:LxTx} plots ICM luminosity (corrected for redshift evolution) against temperature for the results obtained by spectral analysis, and Table~\ref{tab:stats} lists the results of the correlation tests. Since the majority of the temperatures were estimated from a scaling relation (Pratt et al. 2009), it would be startling if there were not a strong correlation for the full data set. The bulk of the scatter comes from the sources with temperatures derived from their spectra, and this is reflected in a weaker correlation for that subpopulation.

Since our data are doubly censored, we used Schmitt's linear regression (Isobe et al. 1986) to obtain the $L_{\rm X}-T_{\rm X}$ relation. This is known to give a biased estimator when there is intrinsic scatter in both variables, so we took the bisector of the two regression lines (Hardcastle \& Worrall 1999).

The solid line in Figure~\ref{fig:LxTx} shows the regression line for all the ERA sources including the upper limts: $L_{\rm X}=6.08^{+1.72}_{-1.34}\times{10}^{44}(T_{\rm X}/5)^{3.12\pm{0.01}}$. The dashed line shows the Pratt et al. $L_{\rm X}-T_{\rm X}$ scaling relation that we used to obtain the estimated temperatures, which is, as expected, very similar.

B\"ohringer et al. (2012) predict a scaling relation slope of 2.70$\pm{0.04}$ and their review of the literature cites slopes of 2.6 to 3.7 for scaling relations derived from observations. Our result is therefore close to B\"ohringer et al.'s model predictions and compatible with existing observational results.

We estimated the total gravitational mass for each cluster via the assumption of hydrostatic equilibrium, using equation (5.113) from Sarazin (1986). We obtained the density gradient from the $\beta$ model parameters (Birkinshaw \& Worrall 1993). We then compared these results with the $M-T$ scaling relation of Arnaud et al. (2005) and found them compatible within the 1$\sigma$ errors.

We also calculated entropy $S$ within $0.1R_{200}$ using $h^{4/3}(z)S=kT/{n_e^{2/3}}$, where $R_{200}$ is the radius at an overdensity of 200 (Arnaud et al. 2005), $kT$ is the ICM temperature and $n_e$ is the electron density. We converted $\beta$ model counts to electron density via the volume emission measure from the {\it apec} model. We compared our results with the scaling relation of Pratt et al. (2010) and again found them within the expected range.

We concluded that, within the experimental errors, the cluster environments of our radio galaxies are comparable to to those of other clusters of similar luminosity, and find no evidence (within our often large uncertainties, and bearing in mind that only ten of the cluster temperatures have been directly measured) that the radio sources have altered the global cluster properties.

\subsection{Comparison with optical measures}\label{sec:opt}

\begin{figure*}
\plottwo{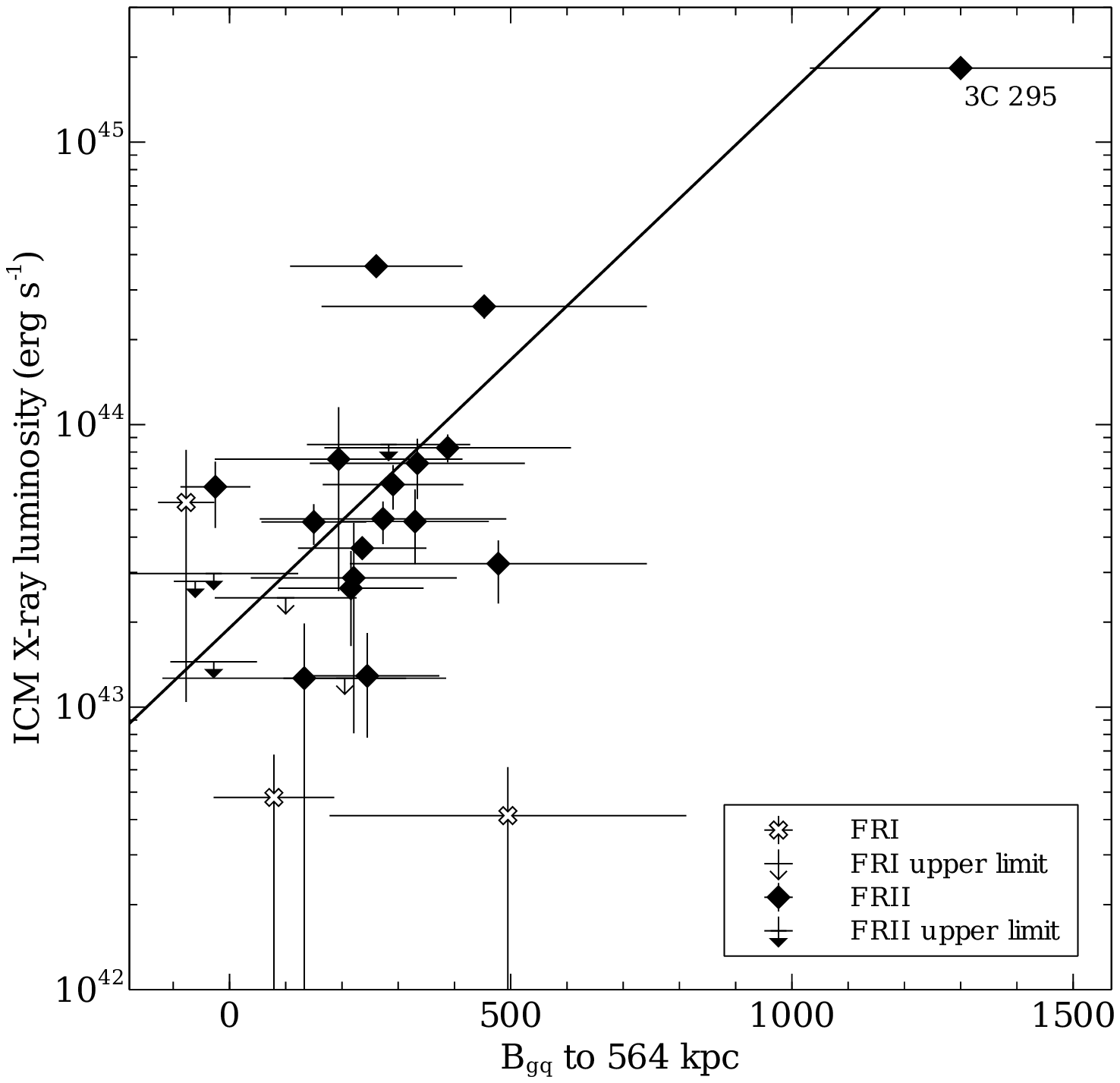}{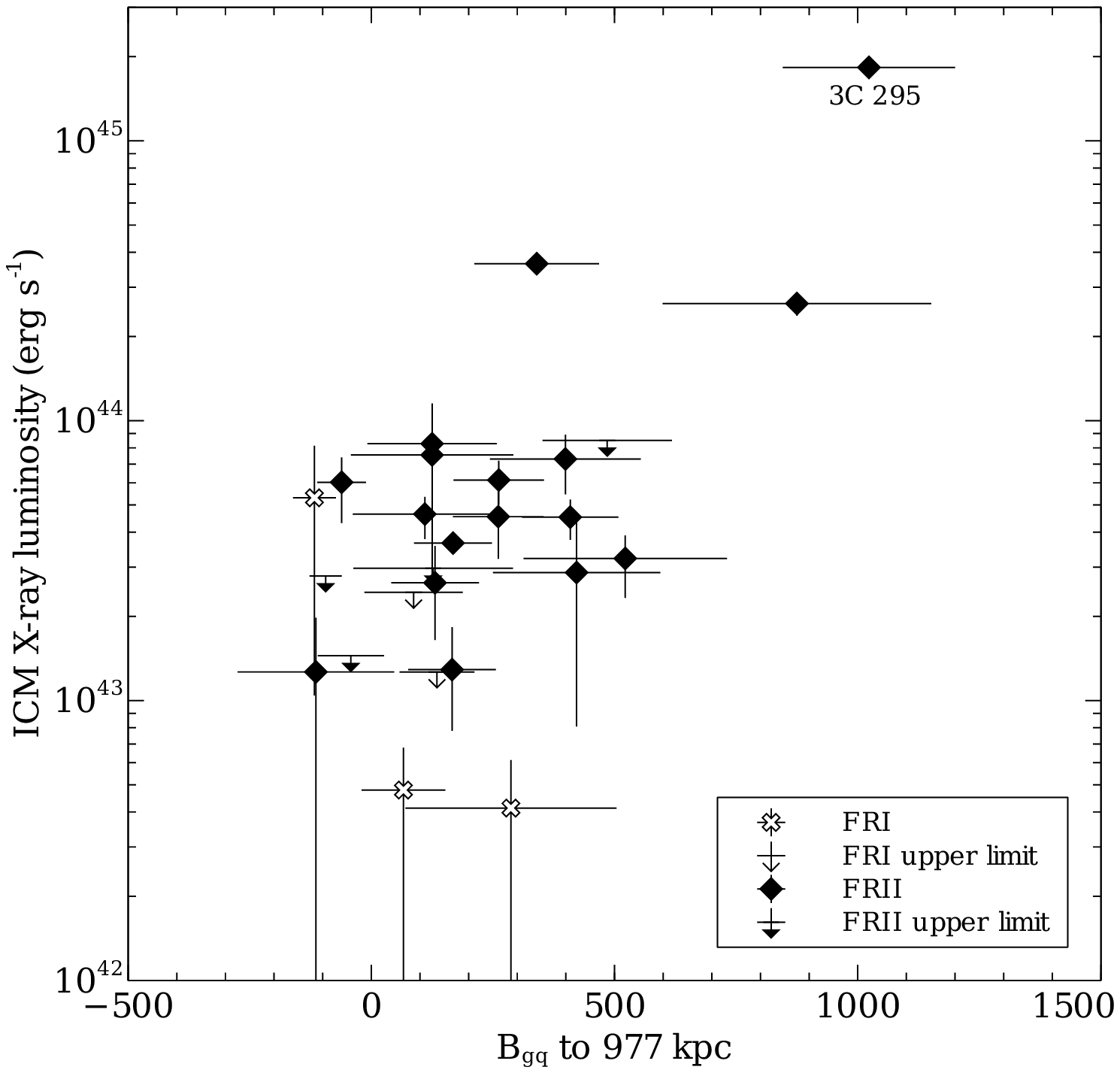}
\caption{ICM luminosity vs the galaxy-quasar spatial covariance function B$_{\rm gq}$ (values taken from Herbert al. 2013). $B_{\rm gq}$ is calculated within 564 kpc (left) and 977 kpc (right). The regression line is calculated for the FRII sources only.}
\label{fig:LxBgq}\end{figure*}

Herbert et al. (2013) have calculated the galaxy-quasar spatial covariance function B$_{\rm gq}$ at two radii (977 kpc and 564 kpc) for the McLure et al. (2004) sample, and we used these values to look for a scaling relation between ICM luminosity and B$_{\rm gq}$ (Figure~\ref{fig:LxBgq}, Table~\ref{tab:stats}). There is only slight evidence of a relationship for the 977~kpc values, but the full 564~kpc data set shows a correlation and this is strengthened when the FRI sources are excluded. Whether this stronger correlation is due to a difference in the two types of galaxy, a change in scaling relation with radio luminosity or some other cause cannot be determined from these data. For the FRII sources, Schmitt's linear regression gives a scaling relation of $\log_{10}(L_{\rm X})=(0.0019\pm{0.0001})\times{B}_{\rm gq}+(43.28\pm{0.11})$, and this is plotted in Figure~\ref{fig:LxBgq}.

\begin{figure}
\plotone{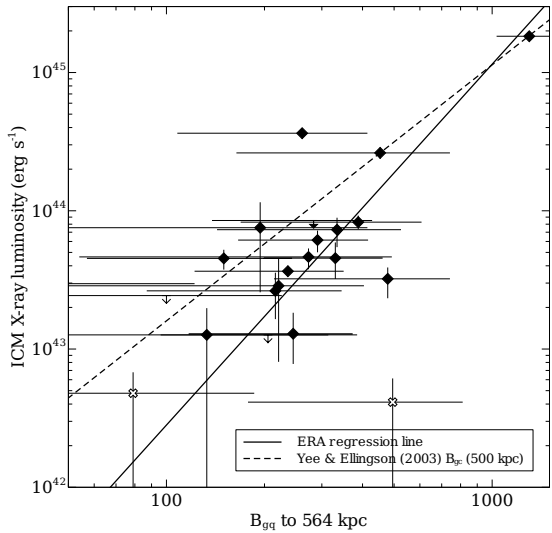}
\caption{ICM luminosity vs $\log_{\rm 10}B_{\rm gq}$ for all positive values of $B_{\rm gq}$ (calculated within 564 kpc) overlaid with the scaling relations from the ERA sample (solid line) and Yee \& Ellingson (2003) (dashed line). Note that Yee \& Ellingson used $B_{\rm gc}$ calculated within 500 kpc.}
\label{fig:LxLBgq}\end{figure}

Yee \& Ellingson (2003) found a power law relation between B$_{\rm gc}$ (the galaxy-cluster center correlation) calculated within 500 kpc, and ICM luminosity -- since their sample is more luminous than the ERA sample, they have no negative values of B$_{\rm gc}$. In order to compare our results with those of Yee and Ellingson, we removed our negative values from the 564~kpc data set and found a good correlation and a scaling relation of $L_{\rm X}=1.71^{+1.14}_{-0.69}\times{10}^{37}(B_{\rm gq})^{2.61\pm{0.01}}$. Figure~\ref{fig:LxLBgq} shows a log plot of the B$_{\rm gq}$ data overlaid with the two scaling relations. Our line is a little steeper than that of Yee \& Ellingson; their sample contains brighter sources over a larger redshift range than ours and their cluster richness measures are calculated within slightly different radii, so a difference in the regression line is not unexpected. 

\subsection{Implications}

Overall, we have found a correlation, significant at the $\sim$~99.5\% confidence level, between radio luminosity and cluster richness. This result is similar to that of Herbert et al. (2013), working with optical overdensity parameters of the McLure et al. (2004) sample, who found a 98\% significant correlation between radio luminosity and environmental overdensity within 564 kpc of the source. Thus for the sample as a whole, the optical and X-ray measures of environment richness yield a similar result.

Splitting the sample by spectral class, we have found a slightly reduced probability of correlation between radio luminosity and cluster richness for LERGs compared with the full data set, but no correlation for HERGs, suggesting that the overall correlation is driven by the LERGs alone. This adds to the body of evidence that there are fundamental differences in the properties of the two classes of radio galaxy. Our HERG clusters occupied a single decade of X-ray luminosity ($10^{43}$ to $10^{44}$ erg~s$^{-1}$), while the LERG clusters had a much wider range of luminosities. However, there is about an order of magnitude of scatter in the relationship, and large uncertainties at the lower end of the luminosity range where most of the upper limits are also situated. This, taken with the small sample size, makes any conclusions tentative.

Best (2004), Hardcastle (2004) and Gendre et al. (2013) all found that HERGs lay within a limited range of relatively low environment richnesses. Hardcastle used $B_{\rm gg}$ within 500~kpc to measure cluster richness. Bearing in mind the weakness of the correlation, the considerable scatter and the measurement radius of 564 kpc rather than 500 kpc, we can use the $L_{\rm X}-B_{\rm gg}$ relationship given in Section~\ref{sec:opt} to estimate that our clusters lie roughly within $-150<B_{\rm gg}<400$, which is roughly compatible with the range occupied by the HERGs in Hardcastle's sample. McLure \& Dunlop (2001) and Harvanek et al. (2001) report similar $B_{\rm gg}$ ranges for their QSOs at $z \sim 0.2$ and $z<0.4$ respectively. For their QSOs in the $0.4<z<0.6$ region, however, Harvanek et al. found higher values of $B_{\rm gg}$, roughly corresponding to cluster luminosities from $10^{43}$ to $10^{45}$ erg~s$^{-1}$.

Looking at results for higher redshift radio galaxies, Belsole et al. (2007) and Wold et al. (2000) both find HERG cluster luminosities within similar ranges. This could hint that any evolution of the HERG environment is slight, but it would be unwise to draw any conclusions based on this cursory analysis. We will look at evolution of the environment in the next phase of the program.

Turning to the LERG subsample, the range of cluster luminosities is larger than for the HERGs, a result also reported by Hardcastle (2004), Best (2004) and Gendre et al. (2013). Best also found a 99.95\% correlation between environment richness and radio luminosity; our correlation is weaker (99.2\%) but supports his result. This suggests a link between radio properties and the environment that does not appear to exist for HERGs. However, when 3C~295 is excluded, the LERG sources are all within about 0.5 dex of the HERG range so the difference between the samples is now slight and the LERG correlation between radio and environment luminosity becomes very weak.

The answer to the question posed in the introduction: whether the
radio-galaxy luminosity is related to large-scale cluster environment
appears to be ``Yes''. There is, however, considerable scatter in the
results, and the correlation is driven by a small number of sources.
We also found tentative evidence that it may be the population of
low-excitation radio galaxies driving this relationship, with no
correlation found for the high-excitation subsample alone. We have
also shown that the correlation between $L_{R}$ and environmental
$L_{X}$ (for the full sample) is not driven by a correlation between
cluster X-ray luminosity and black hole mass.

Such a correlation between radio luminosity and environmental richness
would be expected if jet properties are determined by the properties
of the surrounding hot gas; however, there are many reasons why such a
relationship may be expected to have considerable scatter (as we
observe), even for the narrow redshift range we consider here. An
important source of scatter is the still poorly understood
relationship between jet mechanical power and radio luminosity.
Progress has been made in constraining this relationship
observationally (e.g. B\^{i}rzan et al. 2008, Cavagnolo et al. 2010,
Godfrey \& Shabala 2013), but there is one to two orders of magnitude
scatter. This on its own could be enough to explain the scatter we
observe in the relationship between radio luminosity and environment,
even if jet power and environmental richness were tightly correlated.
It is worth bearing in mind, however, that differences in environment
are the most plausible origin for the scatter in the jet power/radio
luminosity relation, which would act in the opposite direction to
tighten the relationship with environment.

A second important source of scatter is the relationship between
central cooling time and total ICM X-ray luminosity. If jet power is
controlled by the properties of the hot gas environment, then it must
be the central gas distribution that is important. Both cool core and
unrelaxed clusters can be found at all luminosities, and the central
gas density is uncorrelated with total ICM X-ray luminosity for the
cluster population as a whole (e.g. Croston et al. 2008b). It has been
found that low-power radio galaxies appear to require a cool core or
dense galaxy corona (e.g. Hardcastle et al. 2001, 2002a; Sun 2009); however, it has not been observationally established whether
this is the case for the FRII population that form the majority of our
sample. It is therefore plausible that there is substantial scatter
between our measured X-ray luminosities and the central hot gas
properties that may drive jet behaviour (at least in part of the
sample), which could be a major contributor to the scatter we observe
between $L_{R}$ and $L_{X}$.

\section{Conclusions and further work}

We have made a comparison of low frequency radio luminosity and cluster environment richness of a sample of 26 radio-loud AGN. We excluded any effects of environment evolution by taking the sample from a narrow redshift range at $z\sim0.5$. The sample covered three decades of radio luminosity and contained both high- and low-excitation sources. Our measure of environment richness was ICM X-ray luminosity, obtained from {\it Chandra} and {\it XMM-Newton} observations.  

Our main findings are:

\begin{itemize}
\item Over the full sample, there is a correlation between radio luminosity and cluster richness, using ICM luminosity as the measure of cluster richness. There is however total scatter of about one order of magnitude in environment richness at a given radio luminosity, which is not much smaller than the total range of cluster richnesses.
\item There is tentative evidence for a difference between high- and low-excitation sources, with the HERGs occupying a slightly narrower range of cluster richnesses than the LERGs and showing no sign of a correlation between radio luminosity and cluster richness whereas the LERGs have a similar strength correlation to the full sample. However, re-analysis without the brightest source, whose classification as a LERG is disputed, reduces the correlation to only slightly above 95\% significance.
\item We found no evidence for a correlation between radio luminosity and ICM luminosity for the FRII subpopulation.
\item Our results were compatible with published ICM luminosity--temperature scaling relationships. 
\item We compared ICM luminosity and B$_{\rm gq}$ in the hope of finding a scaling relation between the two cluster measures. Although the large scatter makes any such relation dubious, we found a correlation between $\log_{10}L_{\rm X}$ and B$_{\rm gq}$ calculated to 564 kpc (Herbert et al. 2013) for FRII sources, and a power law relationship between $L_{\rm X}$ and B$_{\rm gq}$ that is broadly compatible with that of Yee \& Ellingson (2003). 
\end{itemize}

Having examined a sample of radio-loud AGN at a single epoch, we have found evidence for a correlation between radio luminosity and host cluster X-ray luminosity, as well as tentative evidence that this correlation may be driven by the subpopulation of low-excitation radio galaxies. This is in keeping with previous studies showing different accretion efficiencies and host galaxy properties for the two types of radio galaxy. We also, in common with other researchers, found considerable scatter in the results, which may be a sign of more complex relationships between jet power and environment than are generally assumed.

During the next phase of the program we will compare our results with archive data and published results from different epochs to look for evidence of evolution of the environments of radio-loud AGN.

\acknowledgments

JI and JC acknowledge the support of the South-East Physics Network (SEPnet).

The scientific results reported in this article are based on observations made with the {\it Chandra} X-ray observatory and on observations obtained with {\it XMM-Newton}, an ESA science mission with instruments and contributions directly funded by ESA Member States and NASA. This research has made use of software provided by the {\it Chandra} X-ray Center (CXC) in the application packages \textsc{ciao} and \textsc{chips}, and of the {\it XMM-Newton} Science Analysis Software (\textsc{sas}).

\clearpage

\appendix

\section{Images and surface brightness profiles}\label{app:sb}

This appendix contains images and surface brightness profiles of the ERA sample in order of radio luminosity\footnote{This version of the paper contains a subset of the images and profiles in the full appendix, which will be available in the on-line version of the Astrophysical Journal article. These four examples show profiles with high and low counts for {\it Chandra} and {\it XMM-Newton} observations.}. The images are of the X-ray emission overlaid with radio contours. The dashed circles are the maximum detected radius and the solid circles show R$_{500}$.

Twenty-two sources had sufficient counts to create surface brightness profiles. The PSF and $\beta$ model profiles are overlaid on the profiles. Although profiles were generated for 7C 0223+3415 and 7C 0213+3418, they had insufficient extended emission to fit a $\beta$ model.

\begin{figure*}
\plottwo{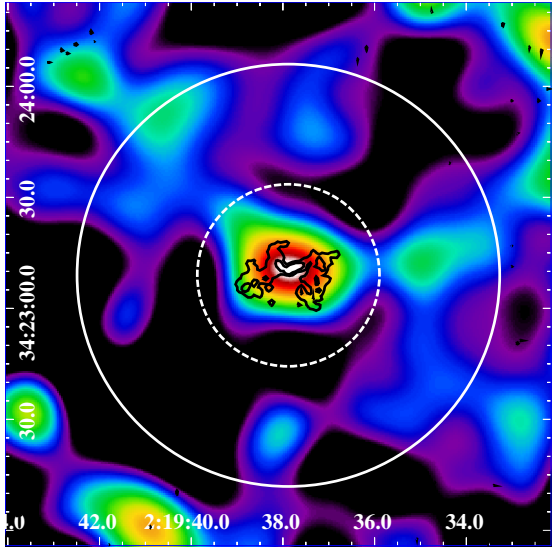}{7C0219_pf_ML.ps}
\caption{7C 0219+3423}\end{figure*}

\begin{figure*}
\plottwo{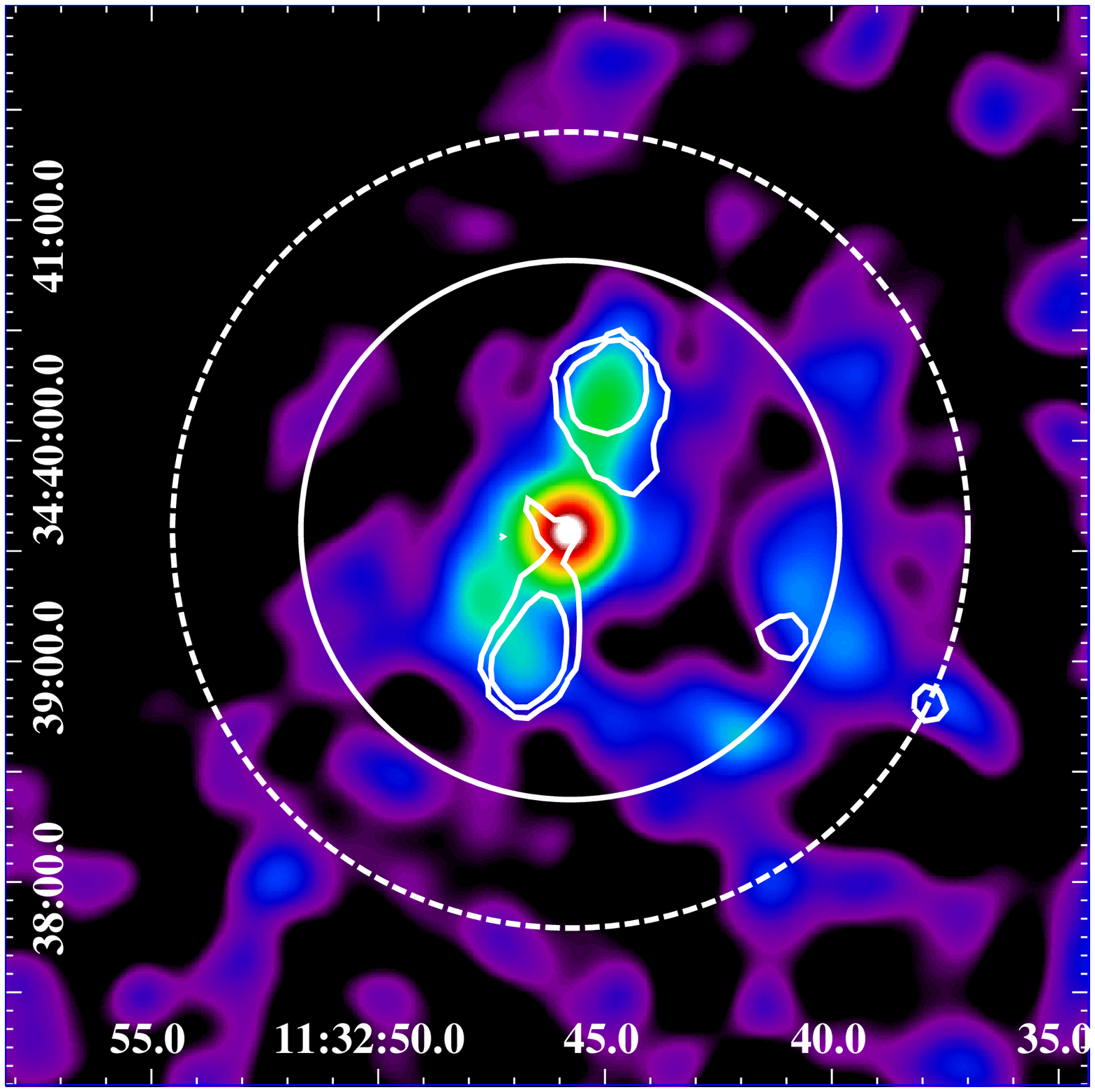}{6C1132_pf_ML.ps}
\caption{6C 1132+3439. The chip edge was excluded for the analysis.}\end{figure*}

\begin{figure*}
\plottwo{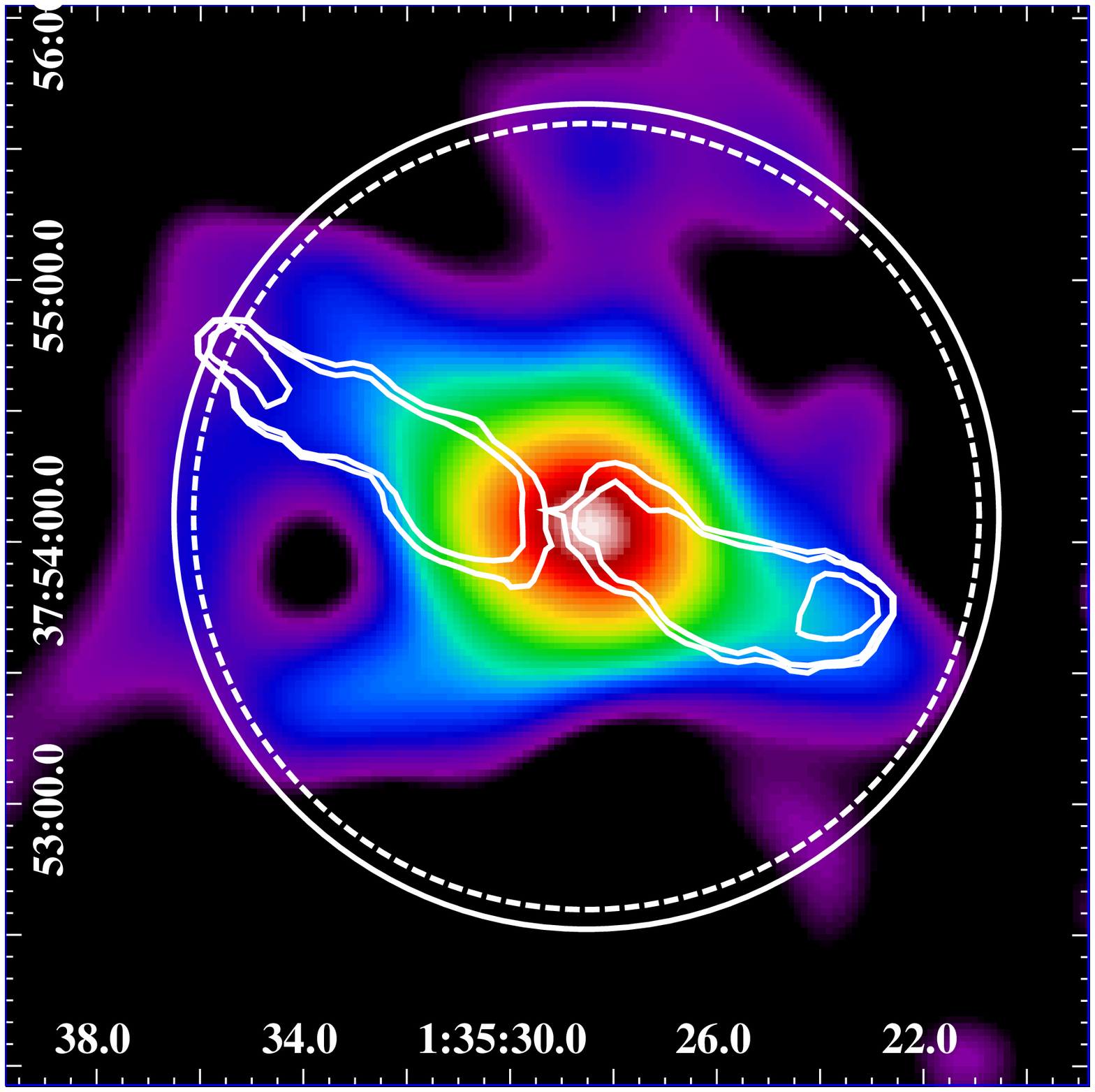}{3C46_pf_ML.ps}
\caption{3C 46}\end{figure*}

\begin{figure*}
\plottwo{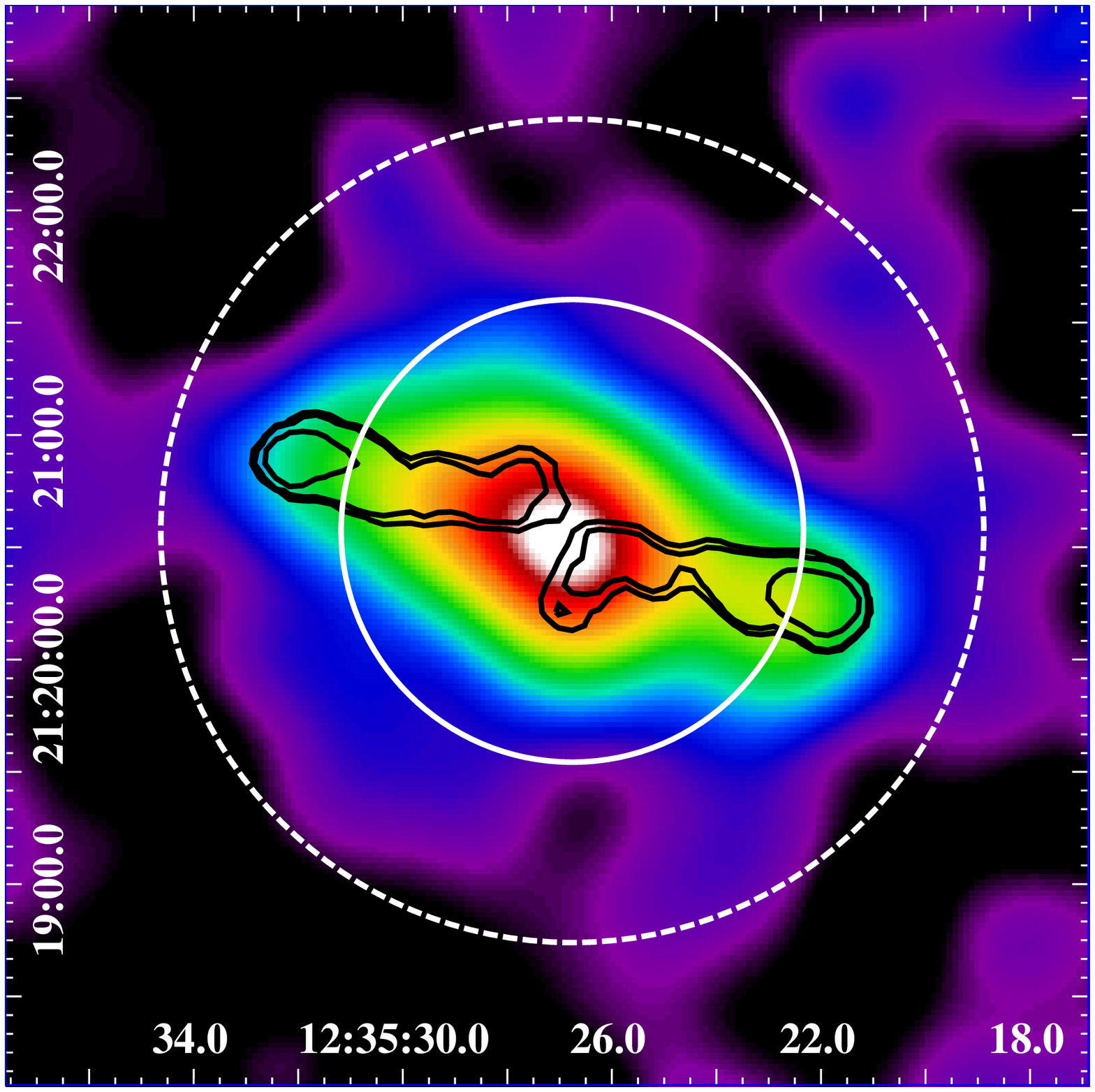}{3C274-1_pf_ML.ps}
\caption{3C 274.1}\end{figure*}

\end{document}